\documentclass[prd,amsmath,amssymb,preprintnumbers,nofootinbib]{revtex4}
\usepackage{aas_macros}
\usepackage{mathrsfs}
\usepackage{xspace}
\usepackage{multirow}
\usepackage{hhline}
\usepackage{array}
\usepackage{adjustbox}
\usepackage{soul}
\usepackage{booktabs}

\newcommand{\beq}{\begin{equation}}
\newcommand{\eeq}{\end{equation}}

\newcommand{\mesa}{{\sc mesa}\xspace}

\newcommand{\heiii}{\ensuremath{{}^3\textnormal{He}}\xspace}

\newcommand{\heiv}{\ensuremath{{}^4\textnormal{He}}\xspace}

\newcommand{\burst}{{\sc burst}\xspace}
\newcommand{\msun}{\ensuremath{M_{\odot}}\xspace}
\newcommand{\np}{\ensuremath{n/p}\xspace}
\newcommand{\yp}{\ensuremath{Y_P}\xspace}
\newcommand{\neff}{\ensuremath{N_{\rm eff}}\xspace}
\newcommand{\gstar}{\ensuremath{g_{\star}}\xspace}
\newcommand{\taun}{\ensuremath{\tau_{n}}\xspace}
\newcommand{\mpl}{\ensuremath{m_{\rm pl}}\xspace}
\newcommand{\tdo}{\ensuremath{T_{D,0}}\xspace}
\newcommand{\nprates}{\ensuremath{n\leftrightarrow p}\xspace rates\xspace}
\newcommand{\bdecay}{\ensuremath{\beta}-decay\xspace}
\newcommand{\tcm}{\ensuremath{T_{\rm cm}}\xspace}
\newcommand{\deltamnp}{\ensuremath{\delta m_{np}}\xspace}

\font\FermiSmallfont=cmssq8 scaled 1200

\def\LANLppthead#1{
	\null
	\begin{center}\vskip -1.0truein{\hbox to 7.0truein {
				\hfill
				\vbox to 1in {\vfill \FermiSmallfont
					\hbox{#1}
					\vfill}
		}}\vskip-0.0truein\end{center}}

\begin{document}
	
\LANLppthead{LA-UR-18-25809}

\title{Nuclear Processes in Other Universes: Varying the Strength of the Weak Force}
\author{Alex R. Howe$^{1}$}
\affiliation{$^{1}$Department of Astronomy, University of Michigan, Ann Arbor,
	Michigan 48109, USA}
\author{Evan Grohs$^{2,3}$}
\affiliation{$^{2}$Department of Physics, University of California Berkeley, Berkeley,
	California 94720, USA}
\affiliation{$^{3}$Theoretical Division, Los Alamos National Laboratory, Los Alamos,
New Mexico 87545, USA}
\author{Fred C. Adams$^{1,4}$}
\affiliation{$^{4}$Department of Physics, University of Michigan, Ann Arbor,
	Michigan 48109, USA}

% Report Number: LA-UR-18-25809

% % % % % % % % % % % % % % % % % % % % % % % % % % % % % % % % % % % % % % % % % % % % % % % % % % % % % % % % % % % % % % % % % % %

\begin{abstract}

Motivated by the possibility that the laws of physics could be different in other regions of space-time, we consider nuclear processes in universes where the weak interaction is either stronger or weaker than observed. We focus on the physics of both Big Bang Nucleosynthesis (BBN) and stellar evolution. For sufficiently ineffective weak interactions, neutrons do not decay during BBN, and the baryon-to-photon ratio $\eta$ must be smaller in order for protons to survive without becoming incorporated into larger nuclei. For stronger weak interactions, neutrons decay before the onset of BBN, and the early universe is left with nearly a pure hydrogen composition. We then consider stellar structure and evolution for the different nuclear compositions resulting from BBN, a wide range of weak force strengths, and the full range of stellar masses for a given universe. We delineate the range of this parameter space that supports working stars, along with a determination of the dominant nuclear reactions over the different regimes. Deuterium burning dominates the energy generation in stars when the weak force is sufficiently weak, whereas proton-proton burning into helium-3 dominates for the regime where the weak force is much stronger than in our universe. Although stars in these universes are somewhat different, they have comparable surface temperatures, luminosities, radii, and lifetimes, so that a wide range of such universes remain potentially habitable.

\end{abstract}

\maketitle

% % % % % % % % % % % % % % % % % % % % % % % % % % % % % % % % % % % % % % % % % % % % % % % % % % % % % % % % % % % % % % % % % % %

\section{Introduction}
\label{sec:introduction}

The laws of physics include a number of fundamental constants with particular values that must be specified, but cannot be derived from currently known theoretical considerations. At the same time, many cosmological models allow for the existence of other universes$-$regions of space-time that trace through independent evolutionary trajectories and are disconnected from our own \cite{rees1997before,Linde17,davies,Donoghue16,Ellis04}. Moreover, the values of the fundamental constants could in principle be different in these other universes. This scenario thus posits a vast ensemble of universes, often called the multiverse, where the various sub-regions sample the different possible versions of the laws of physics \cite{carrrees,carter,davies}. Many authors have suggested that sufficiently large variations in the laws of physics would result in a lifeless universe, so that only small changes to the fundamental constants are allowed \cite{bartip,hogan,reessix}. The goal of this paper is to consider the effects of changing the strength of the weak nuclear force and assess the potential habitability of such scenarios.

Previous work has shown that universes where the weak force is absent entirely could still be habitable for a range of values of the other cosmological parameters and fundamental constants \cite{harnik,paperI}. These papers considered particle physics models and cosmological issues \cite{harnik}, as well as numerical simulations of Big Bang Nucleosynthesis (BBN) and stellar evolution \cite{paperI}. However, previous treatments have not addressed the full implications for universes in which the weak force is weaker than in the Standard Model of particle physics, but still present, or cases where the weak force is stronger. This present paper addresses these more general cases by allowing the strength of the weak nuclear force to vary across the full range of possible values. We focus on nuclear processes, specifically BBN and stellar evolution, and find the strengths of the weak force that allow for universes to be potentially habitable.

The strength of the weak interaction is a fundamental feature of the Standard Model of particle physics (for a textbook treatment see \cite{kane}). The weak interaction determines the rate of beta decay of free neutrons as well as those bound in nuclei. It also controls the rate of helium production in the $pp$-chain for low mass stars because the weak force must act to convert two of the protons into neutrons. Finally, the weak force determines the cross sections for neutrino interactions, which provide an energy drain from stellar interiors and play an important role in the successful detonation of supernova explosions. If the weak force is even weaker than in our universe, neutron decay will be slower and neutrino interactions will be less effective. In the opposite case with stronger weak interactions, neutrons decay rapidly and neutrinos interact with larger cross sections. These processes affect the yields of light elements emerging from BBN and nuclear reactions in stellar interiors. Both of these processes are important for determining the potential habitability of a universe.

For the case of the weakless universe, BBN and stellar evolution play out as follows: the epoch of BBN determines the chemical composition of the universe for the first generation of stars, and has implications for further generations. If nuclear reactions are too effective during BBN, then a universe could process essentially all of its protons and neutrons into heavier elements, leaving no hydrogen behind to make water. In conventional BBN, neutrons start to decay before the onset of BBN, so that protons outnumber the neutrons by a factor of 6$-$7 (depending on when the accounting is done). After essentially all of the neutrons are incorporated into nuclei, mostly helium, this mismatch leads to leftover protons \cite{1990eaun.book.....K}. In the absence of the weak force, however, neutrons do not decay. With equal numbers of protons and neutrons, the universe faces the danger of burning all of its baryons into helium and heavier nuclei. This fate can be avoided if the nuclear reactions cannot proceed to completion during the brief window of time when nucleosynthesis takes place. In particular, lower baryon abundances lead to lower reaction rates, so that even a weakless universe can retain protons. Previous work shows that if the baryon to photon ratio $\eta$ is smaller by a factor of $\sim100$, the helium and hydrogen abundances are the same as in our universe \cite{harnik,paperI}. Other chemical abundances are not the same, as more deuterium and helium-3 are produced, and free neutrons remain.  Later in cosmic history, these nuclear species play an important role in stellar evolution, which is primarily powered by deuterium burning in weakless universes.

With scenarios for both the conventional universe and the weakless universe worked out, this paper considers both the intermediate realm where the weak force is weaker than in our universe, and the opposite case where it is more effective. The weak force has both a strength, set by the Fermi constant $G_F\simeq1.16\times10^{-5}$ GeV$^{-2}\simeq$ (293 GeV)$^{-2}$, and a range, determined by the mass scale of intermediate vector bosons $M_W\sim80$ GeV. In this treatment we keep the masses of all particles the same as in our universe, but we allow the coupling strength and equivalently the Fermi constant to vary, as outlined in Section \ref{constraints}. One important quantity in the problem is the neutron lifetime $\tau_n$, which is mediated by the weak force, and can be written in the form
\begin{equation}
\tau_n^{-1} = {G_F^2 \over 2\pi^3} 
(1 + g_A^2) m_e^5 \lambda_0\,,\label{eq:taun}
\end{equation}
where $g_A\approx1.26$ is the axial-vector coupling for nucleons and  where $\lambda_0\approx1.636$ is a dimensionless parameter. Here we  specify variations in the weak interaction strength in terms of the neutron lifetime.  The limit $\tau_n\to\infty$ corresponds to the weakless universe.

This paper is organized as follows. We first discuss the physical implications of changing the Fermi constant, equivalently the neutron lifetime, in Section II. Using a state of the art numerical code \cite{GFKP-5pts:2014mn}, Section III considers the effects of changing $\tau_n$ on the output from BBN. In this context, we allow the baryon to photon ratio $\eta$ to vary also. Again using a state of the art numerical treatment \cite{mesa}, we consider the effects of changing $\tau_n$ on stellar evolution. In this context, stars of different masses are affected differently, and even the allowed range of stellar masses can vary. Section IV thus considers stellar evolution over the full $(\tau_n,M_\ast)$ parameter space and finds the regions that allow for working stars, as well as the dominant nuclear reactions chains in each regime. In Section V, we examine the later stages of stellar evolution and the chemical evolution and potential habitability of the resulting universes. The paper concludes, in Section VI, with a summary of our results and a discussion of their implications. 

% % % % % % % % % % % % % % % % % % % % % % % % % % % % % % % % % % % % % % % % % % % % % % % % % % % % % % % % % % % % % % % % % % %

\section{Considerations From Fundamental Physics}
\label{constraints}

It is not clear what effect a stronger weak interaction would have on nuclear
structure. Clearly, if the weak force approaches the strength of the strong
force, it ceases to be perturbative, and our models of nuclear structure are no
longer viable. In addition, the nonlinear behavior is complicated by the fact
that the relative strength of the strong and weak forces varies with the energy
scale in question. We can write the weak coupling constant in the low energy
limit in the form 
\begin{equation}
\alpha_{\rm w} = G_F m_P^2 \sim 10^{-5} \,,
\end{equation}
where the neutron lifetime $\tau_n \propto G_F^{-2}$ as given by Equation \eqref{eq:taun}.

Under the conditions in stellar interiors where nuclear reactions take place,
the weak force is less effective than the strong force by $\sim13$ orders of
magnitude. As a result, the neutron lifetime could, in principle, be as short
as $\tau_n\sim10^{-22}\,{\rm s}$ before nuclear reactions are changed
significantly.  As an example, consider the last step in the $pp$-chain where
two \heiii nuclei come together.  The resulting compound nucleus is
beryllium-6, which in our universe disintegrates into \heiv and two free
protons via the strong interaction.  If the neutron lifetime is on order the
strong-interaction timescale, beryllium-6 would decay into lithium-6 at a rate
competitive with the strong branch.  Such a scenario would have profound
implications for main sequence stars.

Note that at less extreme values of $\tau_n$ (equivalently, $G_F$), beta decay (\bdecay)
will occur more readily inside nuclei than in our universe. For all radioactive
nuclei subject to \bdecay, the half-lives will be much shorter. For
sufficiently small values of $\tau_n$, only the stable isotope for a given mass
number $A$ will have an appreciable abundance.  In standard BBN, for example,
the $A=3$ isobar has contributions from both tritium, T, and \heiii.  Much of
the $A=3$ material is eventually synthesized to \heiv through strong and
electromagnetic reactions.  T only has a single proton, and so will
correspondingly have a smaller Coulomb barrier than the $Z=2$ nucleus of
\heiii.  If the weak interaction immediately transmutes T into \heiii, the
rates for \heiv synthesis will decrease which may result in a smaller abundance of
\heiv than the abundance inferred from the neutron-to-proton ratio.
Those ``missing'' neutrons would be in \heiii nuclei which could affect stellar
nucleosynthesis at later epochs.  However, a weak interaction strong enough to transmute T into
\heiii on BBN timescales would also hold the neutron-to-proton ratio at
equilibrium until late times.  As a result, there may be few neutrons 
that survive the BBN epoch and so the relative abundances of \heiii and \heiv
could be unimportant to the first generation of nearly-pure-hydrogen stars.

These factors related to weak versus strong nuclear reactions may be moot,
however, if the strength of the weak force approaches that of the
electromagnetic force.  If the two forces remain unified in an electroweak
force, then the photon and $Z^0$ bosons convert to the $W_3$ and $B_0$
eigenstates, and again nucleosynthesis will not operate normally, particularly
the ${\rm D}(p,\gamma)\heiii$ reaction in the $pp$-chain. At the scale of
nuclear reactions in stars, this unification occurs at $\tau_n\sim 10^{-15}$ s.
It is not clear what form nuclear physics will take in such weakful universes.

The next problem with a weakful universe is that neutrino
interaction cross sections will be nontrivial at stellar and BBN conditions.
Neutrinos will behave more like photons in these environments and will
significantly affect nuclear processes.  In our universe the cumulative optical
depth of neutrinos through the period of BBN is $\sim10^{-12}$. The neutrino
interaction cross section scales inversely with the neutron lifetime, so if we
set $\tau_n\lesssim 10^{-9}\,{\rm s}$, then neutrino interactions will exert a
significant effect on BBN.  Weak nuclear reactions will maintain Nuclear
Statistical Equilibrium (NSE) well into the BBN epoch, and additional
charged-current reactions will be involved, e.g.\,
\begin{align}
	p + p \rightarrow {\rm D} + e^+ + \nu_e,\label{eq:weak_f1}\\
	n + n \rightarrow {\rm D} + e^- + \bar{\nu}_e.\label{eq:weak_f2}
\end{align}
Conversely, high-energy neutrino-spallation can dissociate deuterons through
neutral or charged-current interactions
\begin{align}
	\nu + {\rm D} \rightarrow n + p + \nu,\label{eq:weak_r1}\\
	\nu_e + {\rm D} \rightarrow p + p + e^-,\label{eq:weak_r2}\\
	\bar{\nu}_e + {\rm D} \rightarrow n + n + e^+.\label{eq:weak_r3}
\end{align}
With these reactions occurring, we cannot take our zeroth-order estimate of a
BBN yield of $100\%$ hydrogen for very weakful universes.

For stars, the neutrino opacity limit is even more stringent. The optical depth
of the Sun to neutrinos is $10^{-9}$ \cite{Bahcall95}. A $1\,M_\odot$ star in a
weakful universe would become optically thick to neutrinos at $\tau_n\sim
10^{-6}\,{\rm s}$.  Strong and electromagnetic interactions will still function
normally under such circumstances, but the neutrino bath in the stellar core
will change the progression of stellar evolution, and we can no longer
accurately simulate such objects using stellar evolution codes such as \mesa, as neutrino interactions would
need to be included in its equation of state and nuclear reaction network. (It
is possible to include the effects of neutrino scattering in \mesa, but this is
beyond the scope of the present paper.) Nonetheless, it is entirely possible
that long-lived stars and life-supporting planets could exist in such a
universe.

In addition to neutrino interactions, changes in the weak interaction cross
section will lead to corresponding changes in the abundance of dark matter and
its self-interactions within dark matter halos. Under the usual assumptions for the 
thermal production of weakly interacting dark matter particles (WIMPs), the predicted 
abundance scales as \cite{1990eaun.book.....K}: $\Omega_X \propto \langle\sigma{v}\rangle^{-1} \propto G_F^{-2} m_X^{-2},$
where $m_X$ is the mass of the dark matter particle, and $G_F$ is the Fermi
constant. The usual WIMP miracle is that we expect
new particles (dark matter) to have masses roughly comparable to the weak scale
(100 GeV to 1 TeV) which gives us $\Omega_X$ of order unity. Here we are changing
the value of $G_F$, which would change the expected inventory of dark matter. Since the 
nature of dark matter remains unknown, further discussion is beyond the scope of this paper.

Our analysis so far has focused on weakful universes with shorter neutron
lifetimes.  For longer neutron lifetimes, we maintain the assumption that only
the strong and electromagnetic forces determine binding energies and stability.
A longer neutron lifetime will decrease weak interaction rates at all energy
scales, and we expect convergence to the class of weakless universes in the
limit $\tau_n\rightarrow\infty$.  In this paper, we will take two ranges of
\taun for BBN and stars.  For BBN, we take \taun to be in the range between $1$
and $10^8$ seconds.  If $\taun>10^8\,{\rm s}$, then the weak interaction would
fall out of equilibrium during the quark-hadron transition in the early
universe, which we do not consider quantitatively.  For $\taun<1\,{\rm s}$, we
expect the primordial mass fraction of hydrogen to be nearly equal to unity as
long as nuclear structure remains similar to that for our universe.  We expand
the range of \taun for our stellar calculations to be between $10^{-6}\,{\rm
	s}$ and $10^7\,{\rm s}$, where the lower limit ensures we can still approximate
neutrinos as free-streaming for a $1\msun$ star.  If $\taun>10^7\,{\rm s}$, the
main sequences are virtually identical until a point where the CNO cycle ceases
to be able to power stars.

% % % % % % % % % % % % % % % % % % % % % % % % % % % % % % % % % % % % % % % % % % % % % % % % % % % % % % % % % % % % % % % % % % %

\section{Big Bang Nucleosynthesis with Different Strength of the Weak Interaction}

\subsection{Model}\label{ssec:model}

The weak interaction plays a role in multiple aspects of BBN.  Our motivation
is to investigate how changes to $G_F$ will impact BBN, but there exist other
alternative methods for this pursuit.  Reference \cite{1998PhRvD..57.5480A}
considers what limits primordial and stellar nucleosynthesis can place on the
Higgs mass scale, whereas Refs.\ \cite{2011PhRvC..83d5803B} and
\cite{2014JHEP...12..134H} consider the weak scale in BBN by changing the Higgs
Vacuum Expectation Value (VEV).  Changing the Higgs VEV will change $G_F$ and
the fermion masses of the standard model, assuming that the Yukawa couplings
are preserved while changing the Higgs VEV.  Both Refs.\
\cite{2011PhRvC..83d5803B} and \cite{2014JHEP...12..134H} consider how the
nuclear binding energies would change with different quark masses by using
results from lattice QCD (see references in Ref.\ \cite{2011PhRvC..83d5803B}).
In this work, we will only change $G_F$ in our calculations.  In other words,
we preserve the fermion masses and binding energies by changing the Yukawa
couplings to correspond to changes in $G_F$.

We base our BBN calculations on Refs.\
\cite{WFH:1967,SMK:1993bb,GFKP-5pts:2014mn}.  The standard code from Ref.\
\cite{SMK:1993bb} assumes that the neutrinos are decoupled from the
electromagnetic plasma at all times encompassed by the code (see in particular
Ref.\ \cite{letsgoeu2}).  If we change the strength of the weak interaction by
changing $G_F$, the neutrinos will decouple at either earlier $(G_F<G_{F,0})$
or later $(G_F>G_{F,0})$ times where $G_{F,0}$ is the value of the Fermi
constant in our universe.  We can approximate neutrino decoupling by comparing
the rate of scattering to that of the Hubble expansion rate.  Equations (3) and
(5) in Ref.\ \cite{nu_dec_approx_Hannestad} give expressions for the
annihilation of neutrinos into charged leptons, $\Gamma$, and also that of the
Hubble expansion rate, $H$,
\begin{align}
	\Gamma &= \frac{16G_F^2}{\pi^3}(g_L^2+g_R^2)T^5,\label{eq:weak}\\
	H &= 1.66\gstar^{1/2}\frac{T^2}{\mpl},\label{eq:hub}
\end{align}
where $(g_L^2+g_R^2)$ is the coupling of neutrinos to charged leptons
(dependent on neutrino flavor), \gstar is the effective spin statistic
\cite{1990eaun.book.....K}, \mpl is the Planck mass, and $T$ is the plasma
temperature.  If we take $\gstar=43/4$ and equate Eqs.\ \eqref{eq:weak} and
\eqref{eq:hub}, we can solve for the temperature, $T_D$, at which neutrinos
decouple from the electromagnetic plasma.  Reference
\cite{nu_dec_approx_Hannestad} finds $T_D(\nu_e)\simeq2.4\,{\rm MeV}$ for
electron-flavor neutrinos and $T_D(\nu_\mu,\nu_\tau)\simeq3.7\,{\rm MeV}$ for
$\mu$ and $\tau$ flavor.  For our purposes, it is adequate to use the same
value for all three flavors, which we pick to be $T_{D,0}=3.0\,{\rm MeV}$,
where the $0$ subscript denotes the value in our universe.  We can determine
the scaling of $T_D$ with $G_F$ or the neutron lifetime, \taun, from Eqs.\
\eqref{eq:weak} and \eqref{eq:hub}
\begin{align}
	T_D &= T_{D,0}\left(\frac{G_{F,0}}{G_F}\right)^{2/3}\\
	&= T_{D,0}\left(\frac{\taun}{\tau_{n,0}}\right)^{1/3}.\label{eq:td_tau}
\end{align}
Two problems arise with our scaling law in Eq.\ \eqref{eq:td_tau}.  Firstly,
\tdo is a function of \gstar from Eq.\ \eqref{eq:hub}.  If we increase \taun to
large values, neutrinos will decouple when there are non-negligible amounts of
$\mu$ and $\overline{\mu}$ particles in the electromagnetic plasma.  This complication would
increase \gstar which we did not take into account in Eq.\ \eqref{eq:td_tau}.
Secondly, and related to the first point, if there are $\mu$ particles around,
that would provide more scattering targets for neutrinos.  The result would be an increase in 
the $(g_L^2 + g_R^2)$ factor in a flavor-dependent manner for all three
flavors. The temperature $\tdo\sim[\gstar/(g_L^2 + g_R^2)]^{1/3}$, so the changes in the two
quantities act to offset one another, although the ratio would not be exactly
preserved.  To properly calculate the ratio is non-trivial as $\mu$ and
$\overline{\mu}$ do not fully contribute to either \gstar or the coupling
constants in the temperature range where \tdo may reside.  For simplicity, we
will use the scaling in Eq.\ \eqref{eq:td_tau} to set the decoupling
temperature in our BBN calculations.

Another complication arises if we take a long neutron lifetime.  Neutrons and
protons are formed during the Quark-Hadron transition of $\sim170\, {\rm MeV}$
\cite{2016PrPNP..89...56B}.  Using the scaling in Eq.\ \eqref{eq:td_tau}, the decoupling 
temperature and the Quark-Hadron transition coincide for $\taun\sim10^8\,{\rm s}$.  At this point, neutrinos would
decouple before neutrons and protons exist, and would not directly influence
the neutron-to-proton ratio (denoted \np).  The resulting primordial
nucleosynthesis is much like that of the weakless universe \cite{paperI} with neutrinos acting like dark radiation \cite{dark_rad:2000}.
Therefore, we limit our BBN calculations to $\taun<10^8\,{\rm s}$.

The presence of $\mu$ particles and the modified neutrino decoupling epoch
present two challenges.  Firstly, $\mu$ particles must be added to the
electromagnetic plasma which will change the equation of state.  Equation (D28)
in Ref.\ \cite{letsgoeu2} gives the time derivative of the plasma temperature
during BBN
\beq\label{eq:dtempdt}
\frac{dT}{dt} = -3H\,\left(\frac{d\rho_{\rm pl}}{dT}\right)^{-1}\left(\displaystyle\rho_{\rm pl}+P_{\rm pl}
	+\frac{1}{3H}\frac{dQ}{dt}\biggr|_{T}\right),
\eeq
where $\rho_{\rm pl}$ is the energy density of the plasma (less baryons),
$P_{\rm pl}$ is the pressure exerted by all plasma components, $dQ/dt|_T$ is
the heat lost from the plasma due to nucleosynthesis, and $d\rho_{\rm pl}/dT$
is the temperature derivative of the plasma components (including baryons).  In
Ref.\ \cite{letsgoeu2}, the plasma components are assumed to be photons,
electrons, positrons, and baryons.  We must start our calculation when there
are $\mu$ and $\overline{\mu}$ present, so we must add these terms to the
energy density, pressure, and temperature derivative of energy density.  We
assume that the $\mu$ particles are in thermal equilibrium with zero chemical
potential.  By the time we reach nuclear freeze-out ($T\sim100\,{\rm keV}$),
the equilibrium abundance of $\mu$ particles is $Y_\mu\sim e^{-m_{\mu}/T}$
which is negligible compared to the electron and positron abundances.

Secondly, neutrinos also need to be included in the electroweak plasma equation
of state if they do not decouple until late times.  For temperatures larger
than $T_D$, neutrinos are included in all three plasma terms in Eq.\
\eqref{eq:dtempdt}.  For temperatures lower than $T_D$, they are decoupled and
their effective temperature redshifts with increasing scale factor.  We call
this temperature-like quantity the comoving temperature parameter and denote it
\tcm (see Eq.\ (1) in Ref.\ \cite{xmelec}).

\begin{figure}
	\begin{center}
		\includegraphics[scale=0.60]{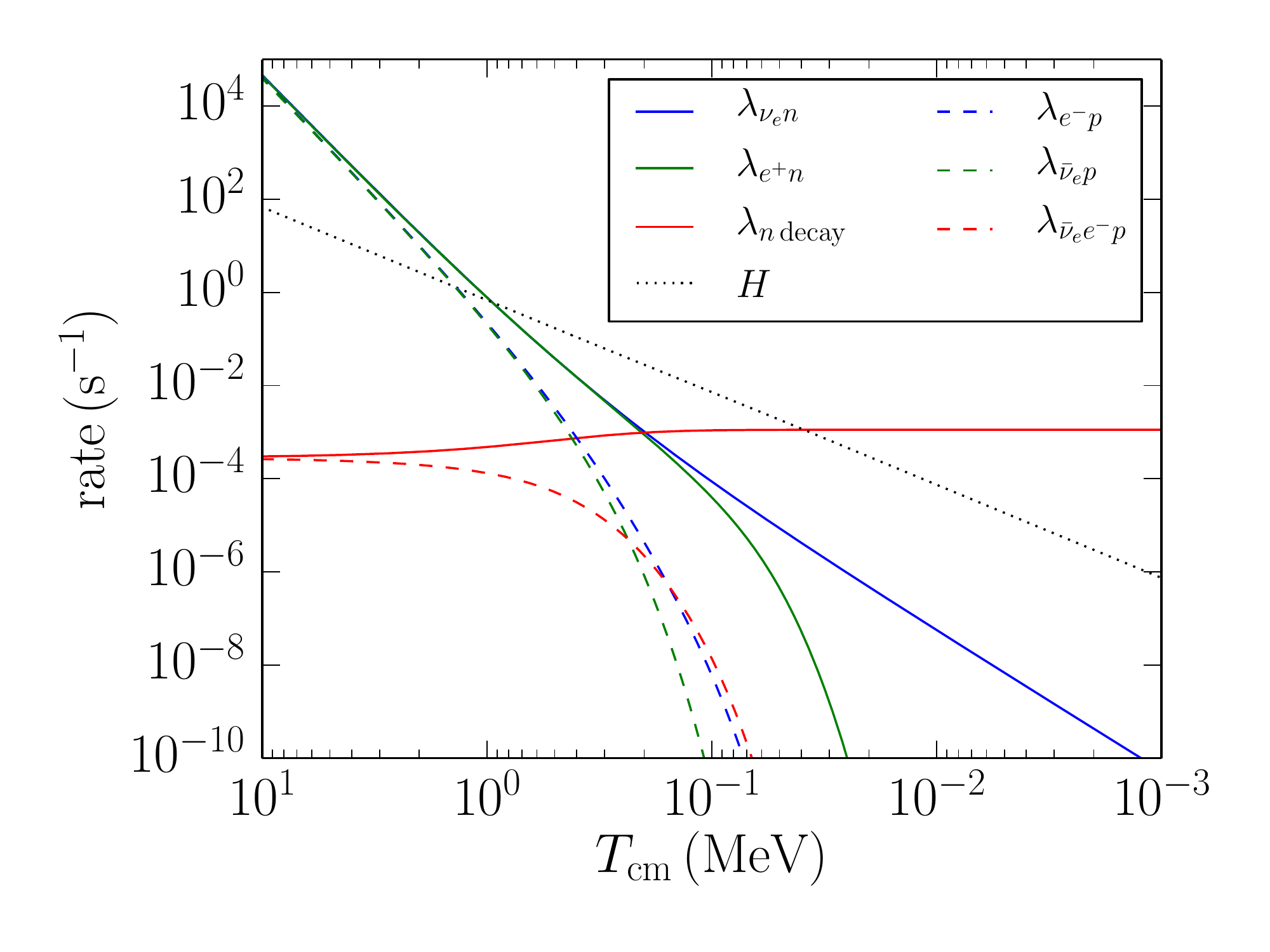}
	\end{center}
	\caption{\label{fig:np_e_v_t} The \nprates in standard BBN versus decreasing
		comoving temperature parameter (equivalent to increasing time) from Ref.\
		\cite{WFO_approx}.  Plotted for comparison is the Hubble expansion rate $H$ as
		a dotted black line.
	}
\end{figure}

The standard nuclear reaction network of Ref.\ \cite{letsgoeu2} contains two
sets of weak rates: the neutron-to-proton rates (denoted \nprates), and the
nuclear \bdecay rates.  The six processes which interconvert neutrons to protons
are given schematically as
\begin{align}
	\nu_e + n &\leftrightarrow p + e^-,\label{eq:np1}\\
	e^+ + n &\leftrightarrow p + \overline{\nu}_e,\label{eq:np2}\\
	n &\leftrightarrow p +\overline{\nu}_e+ e^- .\label{eq:np3}
\end{align}
where the forward arrow indicates proton creation and the reverse arrow
indicates neutron creation.  Each one of the six \nprates [corresponding to a
process in Eqs.\ \eqref{eq:np1} -- \eqref{eq:np3}] scale the same way with
$G_F$ \cite{WFO_approx}
\beq\label{eq:nprate_scaling}
\lambda \propto \frac{G_F^2(1+3g_A^2)}{2\pi^3} I,
\eeq
where $g_A$ is the axial-vector coupling, and $I$ is a phase-space integral
with units of $[{\rm MeV}^5]$ and depends on the specific process in Eqs.\
\eqref{eq:np1} -- \eqref{eq:np3}.  Eq.\ \eqref{eq:nprate_scaling} for
free-neutron decay (the forward process in Eq.\ \eqref{eq:np3}) is consistent
with Eq. \ref{eq:taun} if we take
$I=m_e^5\lambda_0$ at zero temperature.  To wit, we equate the free-neutron
decay rate at zero temperature with the vaccum decay rate, namely $1/\taun$, to
give a quantitative relationship between $G_F$ and the neutron lifetime [see
Eq.\ (26) in Ref.\ \cite{WFO_approx}].  Once we have a value of $G_F$ from an
input value of \taun, we can calculate all six \nprates at all temperatures.
Figure \ref{fig:np_e_v_t} is a plot from Ref.\ \cite{WFO_approx} and gives the
six \nprates of standard BBN as a function of decreasing comoving temperature
parameter \tcm.

If neutrinos decouple from the electromagnetic plasma at high temperatures, it
is possible for there to be other charged-current interactions involving $\mu$
particles, namely 
\begin{align}
	\nu_\mu + n &\leftrightarrow p + \mu^-,\label{eq:npmu1}\\
	\mu^+ + n &\leftrightarrow p + \overline{\nu}_\mu.\label{eq:npmu2}
\end{align}
Processes \eqref{eq:npmu1} and \eqref{eq:npmu2} are the $\mu$-flavor analogs of
\eqref{eq:np1} and \eqref{eq:np2}, respectively.  We do not include a
three-body process with $\mu$-leptons in analogy with the reverse reaction in
process \eqref{eq:np3}, and there is no free-neutron-decay process into a state
with muons.  To calculate the forward and reverse rates for Eqs.\
\eqref{eq:npmu1} and \eqref{eq:npmu2}, we use the same form of the scattering
amplitude as we did for Eqs.\ \eqref{eq:np1} -- \eqref{eq:np3}
\cite{Blaschke_Cirigliano_2016}.  Furthermore, we use the same Coulomb and
zero-temperature corrections for both the $\mu$-flavor and $e$-flavor
\cite{FFN:I,1982PhRvD..26.2694D}.  The key difference between the $\mu$-flavor
and $e$-flavor rates are the mass thresholds.  The neutron-proton mass
difference, \deltamnp, is approximately $1.3\,{\rm MeV}$ whereas the electron
mass is $m_e\sim0.5\,{\rm MeV}$.  As $\deltamnp>m_e$, there are no energy
thresholds for the three forward rates in Eqs.\ \eqref{eq:np1} --
\eqref{eq:np3}, while there are thresholds for the three reverse rates.  The
muon mass is $m_\mu\sim106\,{\rm MeV}$ and so is larger than \deltamnp.
Therefore, the forward process of Eq.\ \eqref{eq:npmu1} does have a threshold
while the reverse process does not.  For Eq.\ \eqref{eq:npmu2}, there is a
threshold for the reverse process and no threshold for the forward process,
similar to the $e$-flavor analog of Eq.\ \eqref{eq:np2}.

The nuclear \bdecay rates of the light nuclei are similar to the \nprates.
Beta decay is a charged-current process with an amplitude which depends on
$G_F^2$.  A difficulty in calculating expressions for \bdecay is that there
exists an amplitude for nuclear transitions which do not have analytic forms.
We will assume that whatever form the nuclear matrix element takes, it is
independent of $G_F$ for the range of values we explore here.  Therefore, we
scale the \bdecay rates the same way we scale the \nprates, namely
$G_F^2\propto1/\taun$.  In practice, our nuclear reaction network employs only
one light-nuclei \bdecay rate, that of Tritium decaying to $^3{\rm He}$.  In
our universe, $\tau_{\rm T}\simeq10\,{\rm years}$, which is much longer than
BBN timescales.  For the tritium decay rate to be important during primordial
nucleosynthesis, \taun would need to be smaller by roughly six orders of
magnitude.

The final BBN ingredient where neutrinos play a role is in the total energy
density.  The Hubble expansion rate is proportional to the square-root of the
total energy density $\rho$.  All of the light-nuclide yields are functions of
the nuclear reaction rates and the Hubble expansion rate.  Helium-4 is
especially sensitive with a faster expansion rate increasing the primordial
mass fraction, denoted \yp.  We adopt the notation of the cosmological
observable \neff to characterize the neutrino energy density
\cite{cmbs4_science_book}
\beq\label{eq:neff}
\rho_{\rm rad}  = \left[2 + \frac{7}{4}\left(\frac{4}{11}\right)^{4/3}
\neff\right]\frac{\pi^2}{30}T^4,
\eeq
where $\rho_{\rm rad}$ is the radiation energy density, and $T$ is the plasma
temperature.  We can relate the neutrino energy density directly to \neff by
taking the radiation energy density as the sum of the neutrino and photon
energy densities
\begin{align}
	\rho_{\rm rad} = \rho_\gamma + \rho_\nu &\implies \rho_\nu
	= \frac{7}{4}\left(\frac{4}{11}\right)^{4/3}\neff\frac{\pi^2}{30}T^4,\\
	&\implies \neff = \frac{4}{7}\left(\frac{11}{4}\right)^{4/3}
	\frac{\rho_\nu}{\frac{\pi^2}{30}T^4}.
\end{align}
In this paper, we always assume the neutrinos have Fermi-Dirac distributions
with temperature parameters equal to the comoving temperature parameter \tcm.
Therefore we can write $\rho_\nu$ as a function of \tcm and solve for \neff in
terms of the ratio $\tcm/T$
\beq
\rho_\nu = 6\frac{7}{8}\frac{\pi^2}{30}\tcm^4\implies\neff
= 3\left[\left(\frac{11}{4}\right)^{1/3}\frac{\tcm}{T}\right]^{4}.
\eeq
To calculate the freeze-out ratio of $\tcm/T$, we use entropy arguments
\cite{1990eaun.book.....K}
\beq
\frac{\tcm}{T}\biggr|_{\rm f.o.} = \left(\frac{g_{\star S,{\rm f.o.}}}
{g_{\star S,{\rm dec}}}\right)^{1/3},
\eeq
where $g_{\star S}$ is the entropic degrees of freedom at a particular epoch.
The f.o.\ subscript denotes the electromagntic freeze-out epoch when photons
are the only plasma particles remaining.  The dec subscript denotes the
neutrino decoupling epoch.  We well consider two limits of neutrino decoupling
to obtain analytic values of the freeze-out ratio of the temperature-like
quantities: early decoupling with muons present; late decoupling with no
electrons present.  In both limits, only photons remain at freeze-out so
$g_{\star S,{\rm f.o.}}=2$.  If muons annihilate after neutrino decoupling
\beq\label{eq:neff_no_mu}
g_{\star S, {\rm dec}} = 9\implies\frac{\tcm}{T}\biggr|_{\rm f.o.}=
\left(\frac{2}{9}\right)^{1/3}\implies\neff\simeq1.56,
\eeq
and if electrons annihilate before decoupling
\beq\label{eq:neff_w_e}
g_{\star S, {\rm dec}} = 2\implies\frac{\tcm}{T}\biggr|_{\rm f.o.}=1
\implies\neff\simeq11.56.
\eeq
If the neutrinos decouple during either $\mu^\pm$ or $e^\pm$-annihilation, then
\neff will be between the two values calculated above.  \neff is close to 3.0
if neutrinos decouple in the interlude between the two annihilation epochs.

%\begin{table*}
%  \begin{center}
%  \begin{tabular}{c c c c c}
%    \hline
%    $\eta$ & & $10^{-11}$ & $10^{-10}$ & $10^{-9}$\\
%    \midrule[1.0pt]
%    $n$ & & $0.2136$ & $3.420\times10^{-2}$ & $3.535\times10^{-3}$ \\
%    \hi & & $0.2139$ & $3.428\times10^{-2}$ & $3.545\times10^{-3}$ \\
%    \hii & & $0.1197$ & $2.615\times10^{-2}$ & $2.766\times10^{-3}$ \\
%    \hiii & & $9.764\times10^{-4}$ & $2.320\times10^{-4}$ & $2.469\times10^{-5}$ \\
%    \heiii & & $8.844\times10^{-10}$ & $2.605\times10^{-10}$ & $2.819\times10^{-11}$ \\
%    \heiv & & $0.4518$ & $0.9051$ & $0.9901$ \\
%    \hline
%  \end{tabular}
%  \end{center}
%  \caption{\label{tab:yields}Mass fractions at different values of $\eta$.
%  }
%\end{table*}

\subsection{Results}

Standard BBN requires the baryon content of the universe as input.  We
adopt the nomenclature of Ref.\ \cite{SMK:1993bb} and use the baryon-to-photon
ratio $\eta$, given as a ratio of baryon and photon number densities
\beq
\eta\equiv\frac{n_b}{n_\gamma}.
\eeq
We do not consider how varying the weak interaction would modify the baryon
content of the universe.  Thermal leptogenesis of Ref.\
\cite{1986PhLB..174...45F} has a dependence on the mass scale in the seesaw
mechanism, and so it is unknown if a different weak scale would change this
particular model of leptogenesis. The ARS model of leptogenesis
\cite{1998PhRvL..81.1359A} is dependent on the Higgs VEV, but also has
dependencies on many other parameters, e.g., Yukawa couplings.  The resulting
baryon asymmetry from ARS leptogenesis may still be unchanged in a large
section of its parameter space.  (See Refs.\ \cite{Davidson:2008bu} and
\cite{Drewes:2017zyw} for reviews on different leptogenesis scenarios.) Our
calculations assume that $\eta$ and $G_F$ are independent of one another and we
will vary both to explore the parameter space. For completeness, note also that 
changing the Higgs VEV can result in instability of the vacuum and could thus affect 
the potential viability of the universe \citep{Sher89}.

\begin{figure}
	\begin{center}
		\includegraphics[scale=0.60]{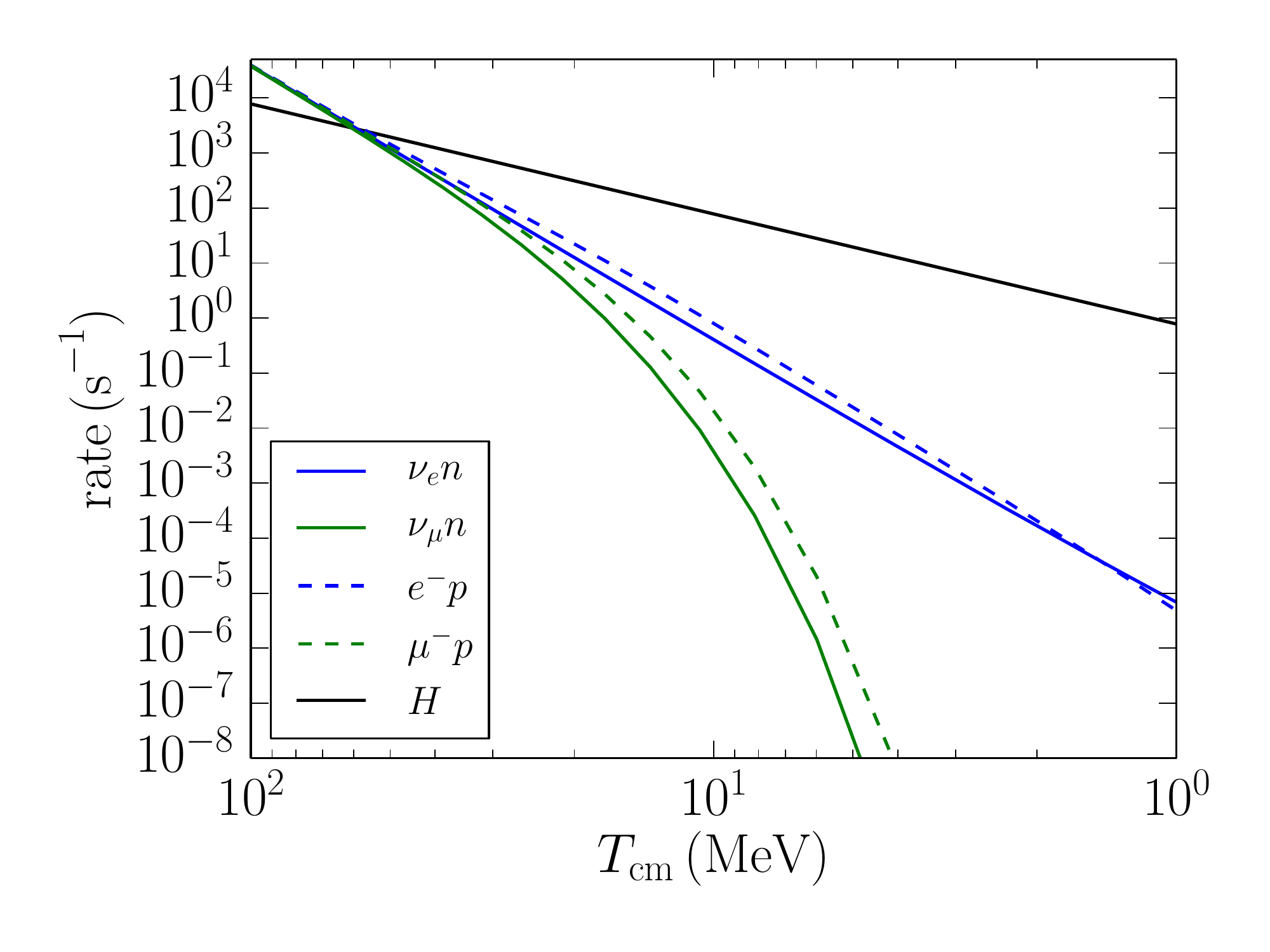}
	\end{center}
	\caption{\label{fig:rate_vs_tcm} Various \nprates and the Hubble expansion
		rate plotted against the comoving temperature parameter, \tcm. The neutron lifetime for this case is
		$\taun=10^8\,{\rm s}$ and the baryon-to-photon ratio is
		$\eta=6\times10^{-10}$.  Blue curves correspond to rates within the
		$e$-type leptons.  Green curves correspond to rates within the $\mu$-type
		leptons.  The solid blue [green] curves correspond to the forward rates of Eq.\
		\eqref{eq:np1} [\eqref{eq:npmu1}], whereas the dashed blue [green] curve
		corresponds to the reverse rate of Eq.\ \eqref{eq:np1} [\eqref{eq:npmu1}].
		Plotted for comparison is the Hubble expansion rate as a dotted black line.
	}
\end{figure}

We first compare the neutron-to-proton rates for $e$ and $\mu$ flavor at large
\taun.  Figure \ref{fig:rate_vs_tcm} shows various \nprates plotted against
\tcm.  We choose $10^8\,{\rm s}$ for the neutron lifetime, and a
baryon-to-photon ratio the same as in our universe, namely
$\eta\simeq6\times10^{-10}$ \cite{PlanckXIII:2015}.  The baryon-to-photon ratio
has little effect on the shape of the curves in Fig.\ \ref{fig:rate_vs_tcm} for
$\eta\ll1$.  The blue curves correspond to rates which involve $e$-type
leptons, namely $\nu_e n\leftrightarrow pe^-$.  The solid curve is the forward
rate (neutron destruction) and the dashed curve is the reverse rate (neutron
production).  The green curves correspond to rates which involve $\mu$-type
leptons, namely $\nu_\mu n\leftrightarrow p \mu^-$, with the same nomenclature
for solid versus dashed linestyle.  Plotted for comparison is the Hubble
expansion rate, $H$, as a solid black line.  Although the \nprates fall below
$H$, \np still evolves on long time scales and so the muon contribution to \np
is non-negligible.  Decreasing the neutron lifetime would increase all four
\nprates (equivalent to shifting the blue and green curves upwards in Fig.\
\ref{fig:rate_vs_tcm}) while preserving $H$.  Freeze-out would occur later and
the contribution from Eqs.\ \eqref{eq:npmu1} and \eqref{eq:npmu2} to \np would
be less significant.  Notice the general trend that all of the \nprates are
roughly equal at high temperature.  At high enough temperature, the three mass
scales of interest (electron, muon, and neutron-proton difference) are either
small or comparable to the temperature.  Therefore, the charged leptons and
neutrinos are relativistic and the amount of available phase-space is the same
for all four rates.  At lower temperatures, the muon mass is important and the
rates involving muon-type leptons become negligible. 

We did not plot the rates from Eq.\ \eqref{eq:npmu2} on Fig.\
\ref{fig:rate_vs_tcm}.  As the muon mass is much larger than the
neutron-proton mass difference, the forward rate of Eq.\ \eqref{eq:npmu2},
namely $\mu^+ n\rightarrow p \overline{\nu}_{\mu}$, is nearly equal to $\mu^-
p\rightarrow n \nu_\mu$.  The two curves are coincident if they had both been
plotted on Fig.\ \ref{fig:rate_vs_tcm}.  The reverse rate of \eqref{eq:npmu2}
suffers from a large threshold in creating the muon mass, so it is always
subsidiary to that of the forward rate.

The mass thresholds in the \nprates of Eqs.\ \eqref{eq:np1} -- \eqref{eq:np3},
\eqref{eq:npmu1}, and \eqref{eq:npmu2} manifest differently for larger \taun
than the value in our universe.  The mass hierarchy is $m_\mu> \deltamnp> m_e$.
When a neutron becomes a proton, there is enough energy to make an electron.
Conversely, a muon incident upon a proton has enough energy to create the
neutron.  As a result, there are thesholds in the reverse rate of Eq.\
\eqref{eq:np1} of the $e$-type.  For the $\mu$-type, there is a threshold in
the forward rate of Eq.\ \eqref{eq:npmu1}.  Figure \ref{fig:rate_vs_tcm} shows
that the dashed green line (no threshold) has a larger rate than the solid
green line (threshold of $m_\mu-\deltamnp$).  What is interesting is that the
blue dashed line (threshold of $\deltamnp - m_e$) is larger than the blue solid
line (no threshold) over multiple Hubble times -- opposite to the behavior of
the $\mu$-type rates and different than our universe (see Fig.\
\ref{fig:np_e_v_t}).  This is primarily due to the fact that the neutrinos
thermally decoupled from the $e^\pm, \mu^\pm$ plasma at a temperature $T_D\sim
100\,{\rm MeV}$.  As the $\mu^\pm$ annihilate, they produce $e^\pm$ pairs and
photons which heat the electromagnetic plasma but not the neutrino seas.  The
$\nu_e$ are at a lower temperature, and so the forward rate of $\nu_e
n\leftrightarrow p e^-$ is smaller than the reverse rate.  Figure
\ref{fig:rate_vs_tcm} shows that the trend eventually reverses at low enough
temperature once the threshold for the reverse rate becomes significant.

\begin{figure}
	\begin{center}
		\includegraphics[scale=0.60]{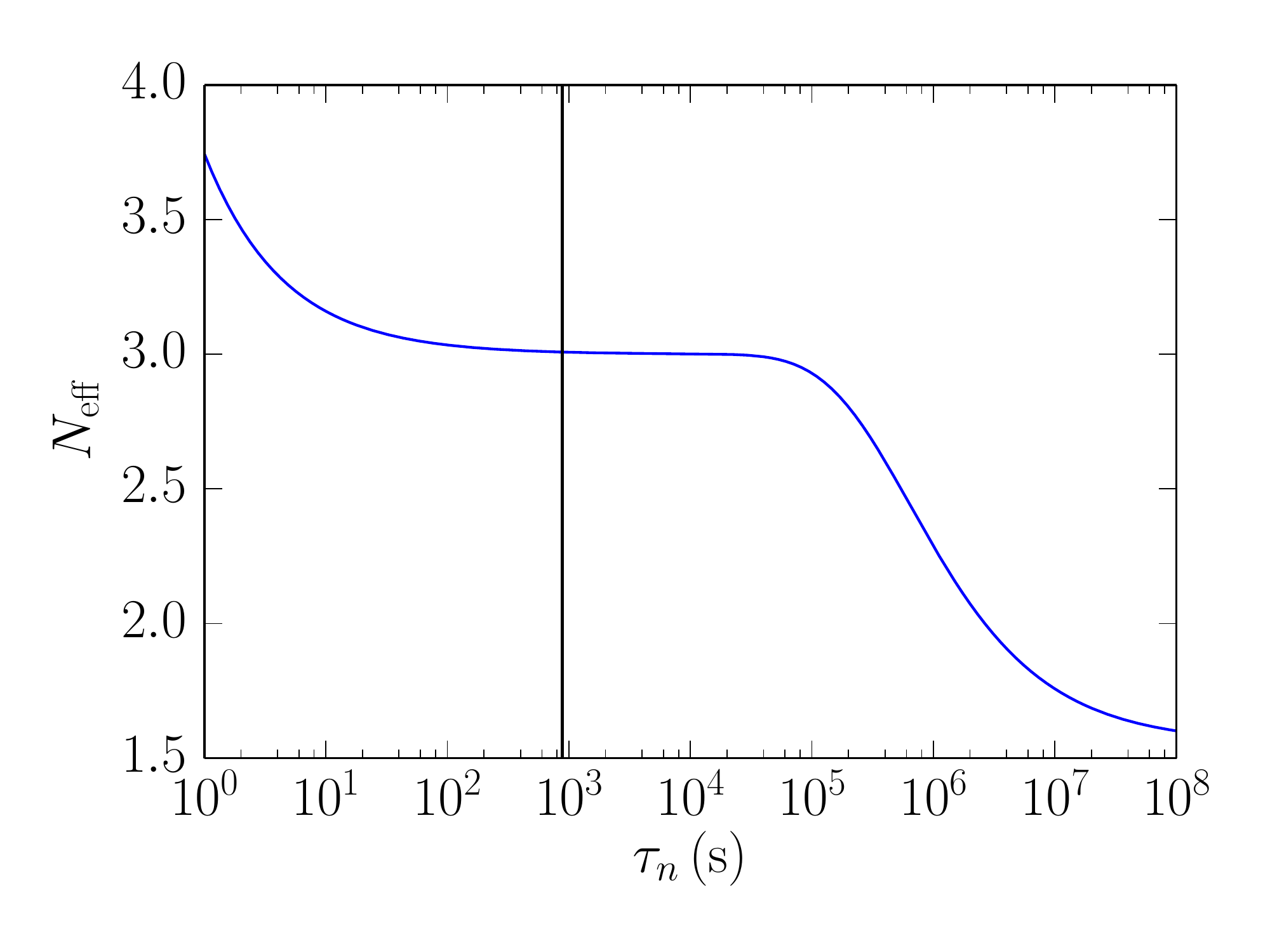}
	\end{center}
	\caption{\label{fig:neff_vs_mntau} Number of effective degrees of freedom $N_{\rm eff}$ 
		(defined in equation [20]) as a function of the mean neutron 
		lifetime $\tau_n$. The baryon to photon ratio is taken to be 
		$\eta=6\times10^{-10}$ for all cases shown here. The black line indicates our universe. Note 
		that $N_{\rm eff}$ is nearly constant over a wide range of 
		$\tau_n$, so that our universe is not fine-tuned in this regard.
	}
\end{figure}

Figure \ref{fig:neff_vs_mntau} gives \neff as a function of \taun for
$\eta=6\times10^{-10}$.  The curve for \neff versus \taun is independent of the
specific value of $\eta$ for $\eta\ll1$.  For large \taun, neutrinos decouple
before the epoch of $\mu^\pm$ annihilation and so \neff is smaller than 3.0.
$\neff$ plateaus to a value of $\simeq1.5$ for large \taun, congruent with the
value in Eq.\ \eqref{eq:neff_no_mu}.  The plateau would not be an asymptote as
even larger \taun would imply earlier decoupling periods when free quarks and
$\tau$ particles are still present.  Our universe sits on the plateau of
$\neff\gtrsim3.0$ in the domain $10\,{\rm s}<\taun<10^5\,{\rm s}$.  We have
assumed that the charged leptons are always given by equilibrium distributions
\beq\label{eq:nl_ng}
n_{\ell}/n_\gamma\propto\left(\frac{m_{\ell}}{T}\right)^{3/2}
e^{-m_{\ell}/T},
\eeq
where $n_\ell$ is the number density and $m_\ell$ the mass of the charged
lepton and we have assumed that we can describe the charged leptons with
Maxwell-Boltzmann statistics -- appropriate for late times when the charged
leptons are annihilating.  The decrease in \neff at $\taun\sim10^5\,{\rm s}$
corresponds to the end of $\mu^{\pm}$-annihilation, whereas the increase in
\neff at $\taun\sim10\,{\rm s}$ corresponds to the beginning of
$e^\pm$-annihilation.  Therefore, the length of this range is a function of the
difference in the electron and muon rest masses, although there is no simple
scaling as we must compare the end of $\mu^{\pm}$-annihilation with the
beginning of $e^\pm$-annihilation and both epochs span multiple Hubble times.
The location of our universe on the plateau informs us that the energy scale of
neutrino decoupling falls between the electron and muon rest masses, as
indicated by the $m_{\ell}/T$ scaling in the exponential of Eq.\
\eqref{eq:nl_ng}.  The final point to make for Fig.\ \ref{fig:neff_vs_mntau} is
that for small \taun, \neff steadily increases above 3.0.  Had we gone to even
smaller values of \taun, we would have seen another plateau at
$\neff\simeq11.5$, consistent with Eq.\ \eqref{eq:neff_w_e}.  Specifically,
$\taun\lesssim3\times10^{-6}\,{\rm s}$ would put neutrino decoupling after
$e^\pm$-annihilation.  In the case of small \taun, the plateua at
$\neff\simeq11.5$ would also be an asymptote as there are no other particles
left to annihilate before photon decoupling.

\begin{figure}
	\begin{center}
		\includegraphics[scale=0.60]{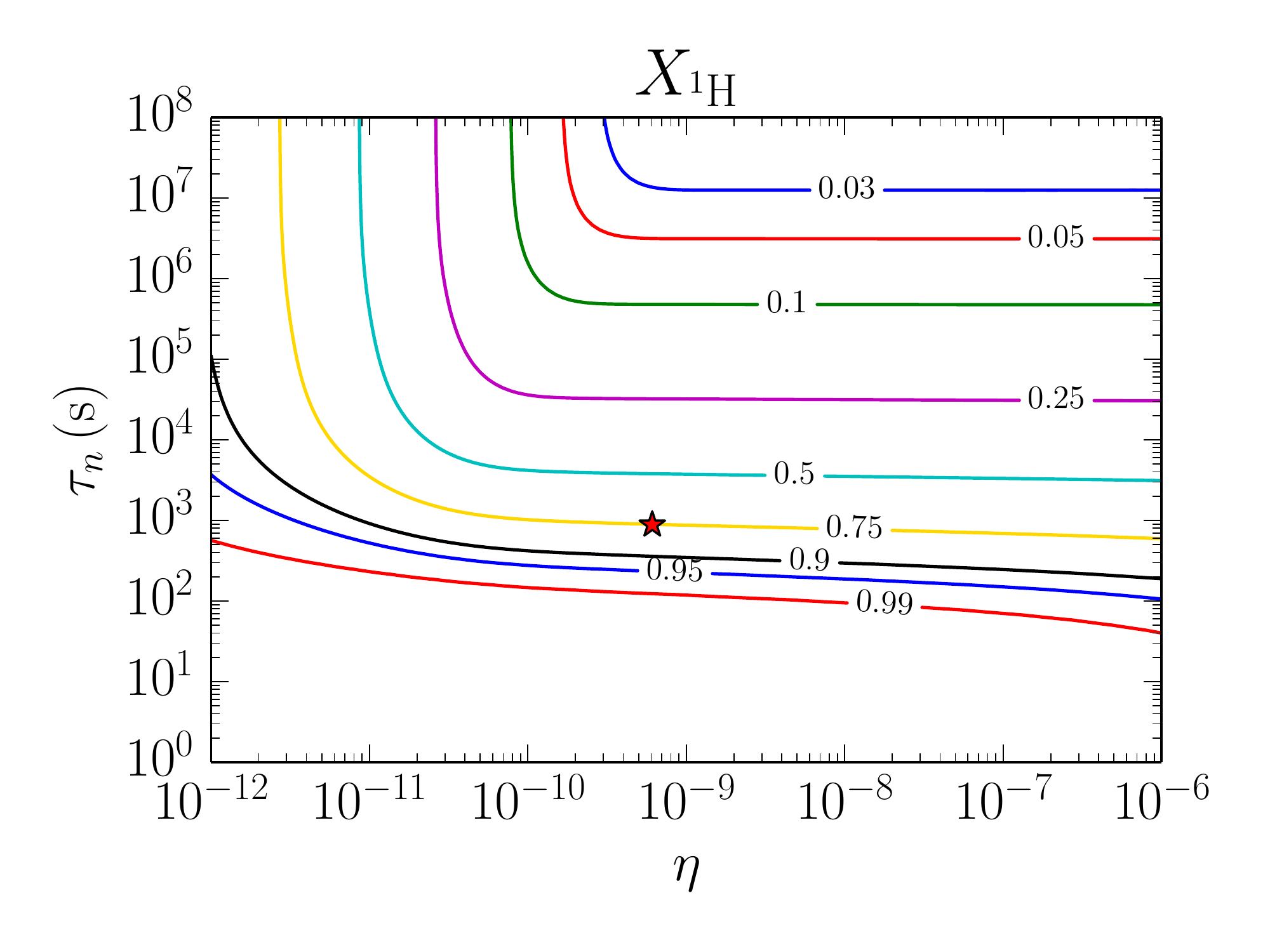}
	\end{center}
	\caption{\label{fig:t_v_e_h1} Contours of constant mass fraction of single-proton 
		hydrogen in the $(\eta,\tau_n)$ plane, where $\eta$ is the 
		baryon to photon ration and $\tau_n$ is the mean neutron 
		lifetime (in seconds). The red star indicates the location 
		of our universe in the diagram. The mass fraction of hydrogan 
		approaches unity in the limit of small $\tau_n$, in the bottom 
		part of the figure, where BBN produces few nuclei. Universes 
		can be rendered sterile in the upper right part of the diagram 
		(large $\eta$ and $\tau_n$), where most of the protons are 
		processed into other nuclei and few are left to supply the 
		universe with water. 
	}
\end{figure}

Figures \ref{fig:rate_vs_tcm} and \ref{fig:neff_vs_mntau} both depict neutrino
physics in the early universe.  The neutrinos change \np which eventually
affects the primordial abundances.  Figure \ref{fig:t_v_e_h1} shows \taun
versus $\eta$ at contours of constant single-proton hydrogen mass fraction,
$X_{^1{\rm H}}$.  The value of $\eta$ is the final baryon-to-photon ratio
after $e^\pm$-annihilation, consistent with what an observer would measure in
the cosmic microwave background.  We place a red star at the value of $\eta$
and \taun which correspond to our universe.  The red star falls on the $75\%$
contour, in line with standard BBN calculations.  For small \taun, the contours
steadily approach an asymptote of unity as few neutrons survive the BBN epoch.
There exist two general trends: increasing $\eta$ gives fewer free protons; and
increasing \taun also gives fewer free protons.  The later trend is consistent
with the scaling that a weaker weak force allows \np to go out-of-equilibrium
at earlier times and hence larger values.  Fewer total protons implies fewer free
protons survive nuclear freeze-out.  A larger $\eta$ implies faster nuclear
reactions and delays nuclear freeze-out to later times.  This trend is present
for small to intermediate values of $\eta$ in Fig.\ \ref{fig:t_v_e_h1}.  For
larger values of $\eta$, the nuclear reactions incorporate the free neutrons
into heavier nuclides and the free-proton mass fraction is stagnant with
increasing $\eta$.

\begin{figure}
	\begin{center}
		\includegraphics[scale=0.60]{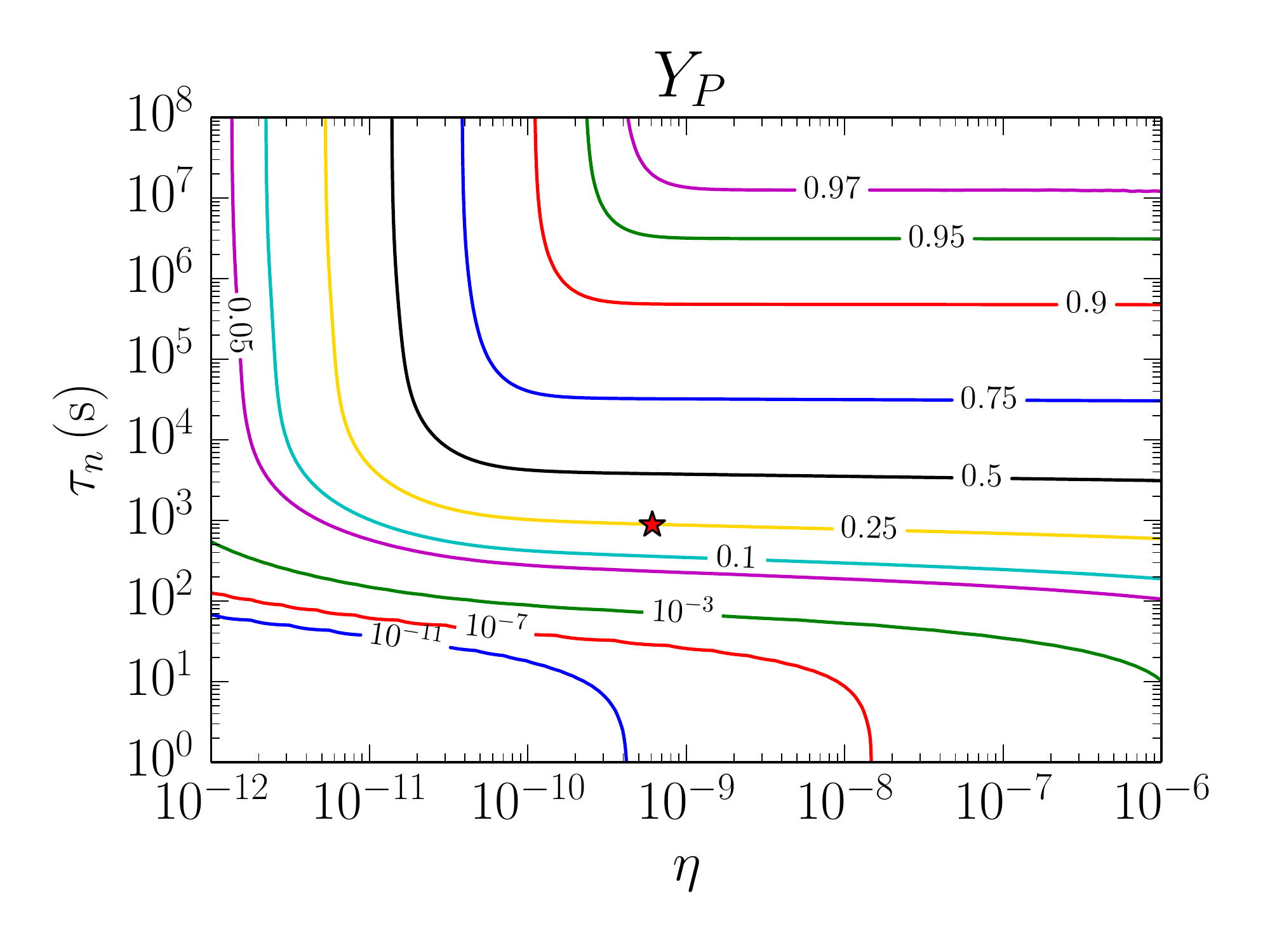}
	\end{center}
	\caption{\label{fig:t_v_e_he4} Contours of constant mass fraction of helium-4
		in the $(\eta,\tau_n)$ plane, where $\eta$ is the baryon to 
		photon ration and $\tau_n$ is the mean neutron lifetime (in 
		seconds). The red star indicates the location of our universe 
		in the diagram. Little helium is produced during BBN for the 
		lower left part of the diagram (small $\eta$ and $\tau_n$). 
		In contrast, the mass fraction of helium becomes large (and 
		hence probletmatic) in the upper right part of the figure 
		(large $\eta$ and $\tau_n$). 
	}
\end{figure}

For large $\eta$, we can make a stronger statement than a constant $X_{^1{\rm
H}}$ and say \np is constant with large $\eta$.  Figure \ref{fig:t_v_e_he4}
shows contours of constant $^4{\rm He}$ mass fraction in the $\eta$
versus \taun parameter space.  The contours in the top-right of the parameter
space (at large $\eta$ and large $\taun$) are equal to $1-X_{^1{\rm H}}$ to
high precision.  In addition for small \taun, there are very few neutrons which
survive weak-freeze-out and little helium is produced.  The red star on Fig.\
\ref{fig:t_v_e_he4} denotes the values of $\eta$ and \taun in our universe and
lies on the $25\%$ contour.  We can see that on both Figs.\ \ref{fig:t_v_e_h1}
and \ref{fig:t_v_e_he4}, the contours are flat horizontally and regularly-spaced
vertically in the local area of each red star.  The standard formula
\cite{1990eaun.book.....K} relating \np to helium and hydrogen holds in this
local vicinity
\begin{align}
	\yp &= \frac{2\np}{1+\np}\label{eq:yp_np1},\\
	X_{^1{\rm H}} &= 1 - \yp,\label{eq:yp_np2}
\end{align}
where \np is a function of \taun and independent of $\eta$.

\begin{figure}
	\begin{center}
		\includegraphics[scale=0.60]{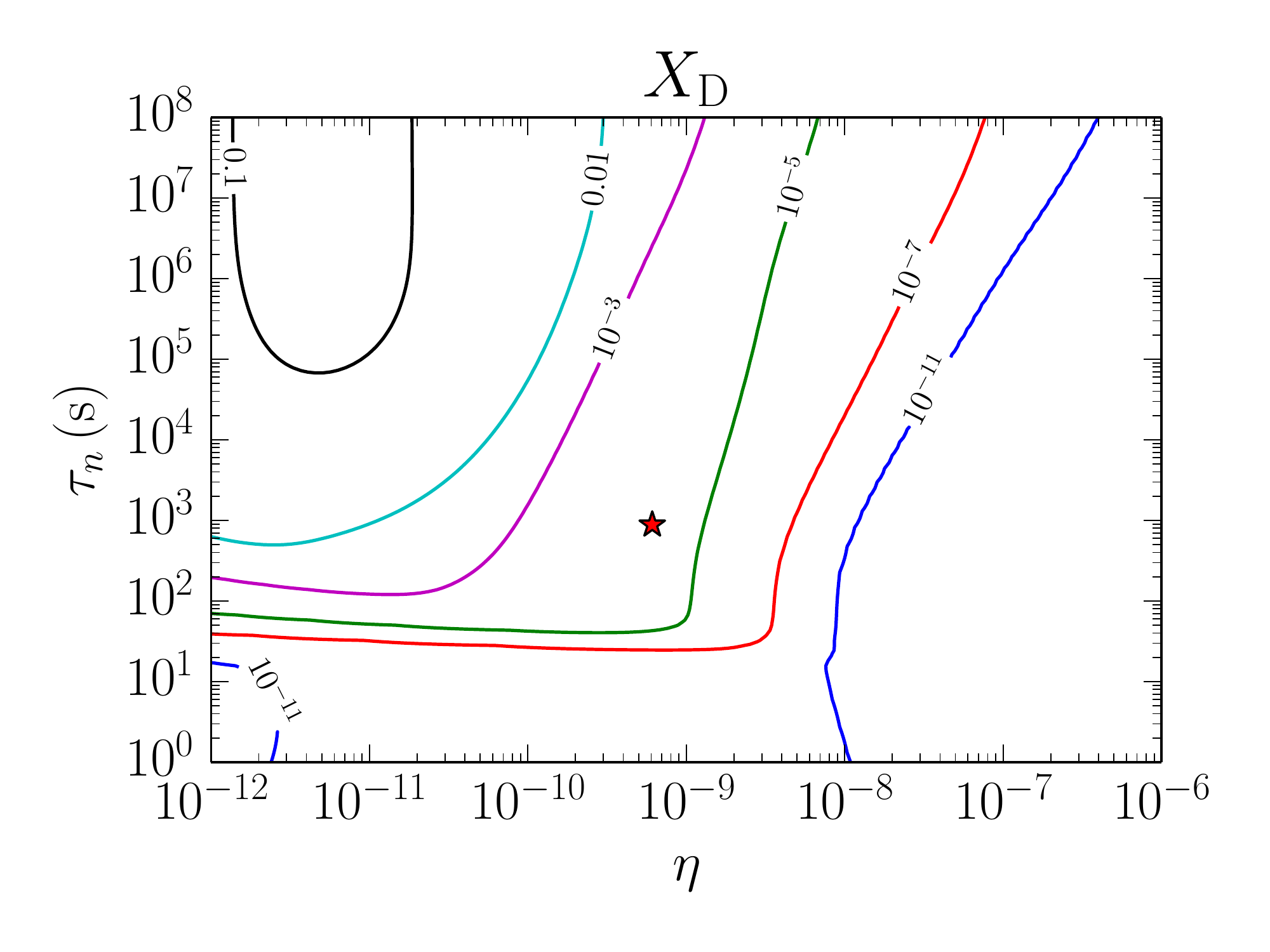}
	\end{center}
	\caption{\label{fig:t_v_e_d} Contours of constant mass fraction of deuterium 
		in the $(\eta,\tau_n)$ plane, where $\eta$ is the baryon to 
		photon ration and $\tau_n$ is the mean neutron lifetime (in 
		seconds). The red star indicates the location of our universe 
		in the diagram. Only trace amounts of deuterium are produced 
		by BBN over the entire plane, except for the upper left part 
		of the figure where $\tau_n$ is large and $\eta$ is small. 
	}
\end{figure}

In fact, the relationships in Eqs.\ \eqref{eq:yp_np1} and \eqref{eq:yp_np2} hold
over the entire parameter space shown in each figure except for the small
$\eta$ and large $\taun$ quadrant.  The \np ratio is large in this area, but
the reaction rates are slow and so \yp is underproduced with respect to the
relation in Eq.\ \eqref{eq:yp_np1}.  Figure \ref{fig:t_v_e_d} shows contours of
constant deuterium mass fraction in the \taun versus $\eta$ plane.  In the
upper-left quadrant, we see that $X_{\rm D}$ reaches the $10\%$ level, roughly
four orders of magnitude larger than the calculation using the values of $\eta$
and \taun from our universe.  \np is stil largely preserved as a deuteron has
the same ratio of neutrons to protons as a \heiv nucleus.  One issue regarding
deuterium which may be relevant for habitability is the abundance of water.  We
do not know the minimum hydrogen abundance for life-sustaining water, or if
there is such a nonzero minimum. Reference \cite{paperI} argues that a
weakless universe with a low value of $\eta$ and $\np\simeq1$ does not preclude
abundant water due to a relatively large abundance of deuterium.  In the weaker
universes in the upper-left quadrant of Fig.\ \ref{fig:t_v_e_d}, $X_{^1{\rm
H}}$ is larger than $X_{\rm D}$, so there is no problem with a lack of hydrogen
in this area of the parameter space.  Fig.\ \ref{fig:t_v_e_h1} shows that the
upper-right quadrant of the parameter space suffers from a small $X_{^1{\rm
H}}$, which is coincident with small $X_{\rm D}$ in Fig.\ \ref{fig:t_v_e_d}.
Therefore, deuterium is not a solution to the problem of a lack of hydrogen for
water in a weaker universe.

We have only included the \bdecay rates and \nprates in the weak sector of our
nuclear-reaction network, but we can give a qualitative description of what
could occur in the presence of other weak interaction rates.  In a weakful
universe, neutrinos can interact with nuclei through neutral and charged
current processes which could yield rates comparable to the strong and
electromagnetic nuclear interactions. (These are not included in the current code.) We included examples of weak
interactions with deuterons in Eqs.\ \eqref{eq:weak_f1} -- \eqref{eq:weak_r3}.
Nuclear excited states begin a few MeV above the ground.  At temperatures
comparable to the nuclear excited states, the strong and EM nuclear rates are
fast enough to keep the lightest nuclei in NSE.  We would expect that enhanced
weak interactions would keep the nuclear abundances in NSE to lower
temperatures.  This would not necessarily cause a decrease in the {\it total}
\np ratio as nuclear reactions would supplement the \nprates in keeping \np in
equilibrium.  For example, the neutrino charged-current interaction with
deuterons, Eq.\ \eqref{eq:weak_r2}, gives a secular abundance of
\begin{align}
	Y_{\rm D}&\sim Y_p^2\frac{n_b}{(mT)^{3/2}}e^{(2m_p-m_{\rm D})/T}\\
	&\sim Y_p^2\eta\left(\frac{T}{m}\right)^{3/2}e^{(2m_p-m_{\rm D})/T},
\end{align}
where $Y_p$ is the free proton abundance and $m$ is the baryon mass.  If we
assume $Y_p$ is still of order unity, we see that we can have a significant
deuterium abundance at low temperature as $2m_p>m_{\rm D}$.  If this is the
case, then BBN would result in significant production of deuterium and perhaps 
heavier nuclei at small \taun rather than resulting in nearly all $^1$H. We stress 
that this is an NSE expression and therefore a rough approximate, as under these 
conditions, the universe will not be in NSE.

% % % % % % % % % % % % % % % % % % % % % % % % % % % % % % % % % % % % % % % % % % % % % % % % % % % % % % % % % % % % % % % % % % %

\section{Weaker and Weakful Stars}

We computed evolutionary models of weaker and weakful stars with the stellar evolution code \mesa. We created nuclear reaction networks that reflected a different strength of the weak force by multiplying the $pp$ and pep reaction rates by the inverse of the ratio of the neutron lifetime to that in our universe ($\tau_{n,0}/\tau_n$). Even at the smallest value of $\tau_n$ we consider, the $\rm{D}(p,\gamma)^3$He reaction is fast compared with the $p(p,e^+\nu)\rm{D}$ reaction (seconds versus days), so we retain \mesa's treatment of the two as a single reaction where every $pp$-reaction is immediately followed by a D-$p$ reaction to produce $^3$He. For sufficiently large $\tau_n$, we make an additional change to reflect the fact that beta decays in the nuclear reaction chains will be longer than the lifetime of the star. In this situation, we delete the CNO reactions and the beta decays of $^7$Be and $^8$B from the nuclear reaction network. The nuclear reactions in later stages of stellar evolution are all strong reactions, so they do not need to be changed.

In the results of our simulations, weaker and weakful stars can be characterized by the dominant energy-producing process powering them. This process varies with the neutron lifetime, stellar mass, and composition, but follows some general patterns. In this section, we consider four general cases: first zero-metallicity (Population III) and solar-metallicity (Population I) stars at constant $\eta$, holding the cosmological parameters constant. Then, we consider zero-metallicuty and solar-metallicity at a constant $^4$He fraction, effectively adjusting $\eta$ to keep the $^4$He fraction produced by BBN as nearly constant as possible. A constant $Y_P$ combined with a small $\tau_n$ requires a value of $\eta$ larger than the range we explore, and for a suffiently small $\tau_n$, it may be unphysical. Nonetheless, we examine this scenario as a useful point of reference to give a fuller picture of the nuclear processes at work. Stars in small-$\tau_n$ universes would likely form with a near 100\% $^1$H composition and would have a lifetime $\sim$33\% longer than the ones we model. We consider a range of $\tau_n$ ranging from the age of the universe (at which point free neutrons survive to the epoch of star formation, and the scenario is functionally equivalent to the weakless universe) down to the limit of $10^{-8}$ to $10^{-4}$ s, at which our models of stellar evolution break down. We also note the point at which our models of BBN break down.

\begin{figure}[ht]
	\includegraphics[width=0.99\columnwidth]{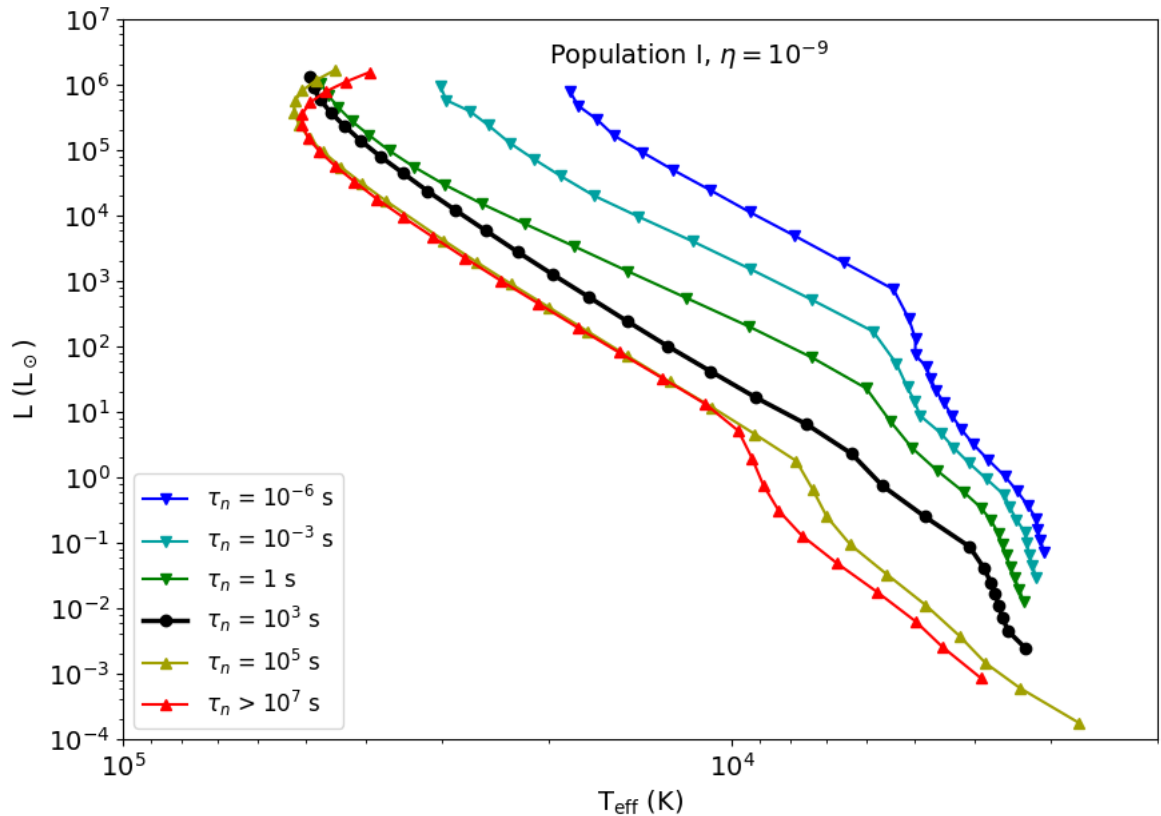}
	\caption{H-R diagram showing the stellar main sequence for 
universes with different strengths of the weak interaction. 
The curves show the main-sequence for universes with neutron 
lifetimes ranging from $10^{-6}$ s (blue) to $10^7$ s (red). 
The stellar masses range from 100 $M_\odot$ down to the 
minimum mass appropriate for the given universe (as indicated 
in Figure 9). Universes with neutron lifetimes longer than 
$10^7$ s have essentially identical main-sequences, which are that of the 
weakless limit. Universes with lifetimes shorter than $10^{−6}$ s 
involve new nuclear processes relating to neutrino interactions, 
and cannot be modeled with existing stellar evolution codes.
All stellar models use the starting composition for Population I 
stars in universes with $\eta=10^{-9}$ (see text). }
	\label{HRplot}
\end{figure}

Figure \ref{HRplot} shows the main sequence for stars over the range of $\tau_n$ we study, from $10^{-6}$ s to $10^7$ s. The latter is virtually identical to the main sequence at long $\tau_n$ out to the point where the CNO cycle shuts down because this is the dominant process in this range (see Section \ref{dominant}). In this figure, we plot only Population I stars in the constant-$\eta$ case, because the main sequence does not look significantly different in the other cases, except for being bluer for Population III, as expected. In other words, stars look fairly similar over a wide range of $\eta$ and $\tau_n$ as long as they actually attain core hydrogen burning, which is the longest stage of their life cycle and thus comprises the main sequence over most of the parameter space we study. A longer neutron lifetime (and a weaker weak force) makes the main sequence bluer, but with a similar shape. A shorter neutron lifetime makes the main sequence both redder and steeper, approaching the Hayashi track, where more efficient strong-burning stars fall \citep{paperI}. A detailed study of the effects of varying the strength of the weak force is given in the following subsections.

\subsection{Dominant Nuclear Processes}
\label{dominant}

This parameter study also requires a clear definition of what a star is. In some parts of the parameter space, ``stars'' will form with a large fraction of deuterium. This means that deuterium burning will be much more important than in our universe, and long-lived deuterium-burn objects could exist. This raises the question of how to distinguish a star from a brown dwarf in such a case. For the purpose of this paper, we define a deuterium-burning object to be a star if it has a composition of at least 2\% deuterium by mass, for two reasons. First, it is approximately the minimum deuterium fraction required for a minimum-mass star to have a main-sequence lifetime of 1 Gyr, and thus it is the lowest deuterium fraction that is important for habitability purposes. Second, as $\tau_n$ increases to infinity with $\eta$ remaining the same as in our universe, the limiting fraction of light hydrogen is also $\sim$2\%, which provides a nice symmetry with our deuterium fraction limit. A deuterium fraction of $\geq 2\%$ never occurs in either of our constant-$\eta$ cases (see Figure \ref{fig:t_v_e_d}), but it occurs in both of our constant-$Y_P$ cases for $\tau_n \geq 3000$ s.

\begin{figure*}[ht]
	\includegraphics[width=0.99\textwidth]{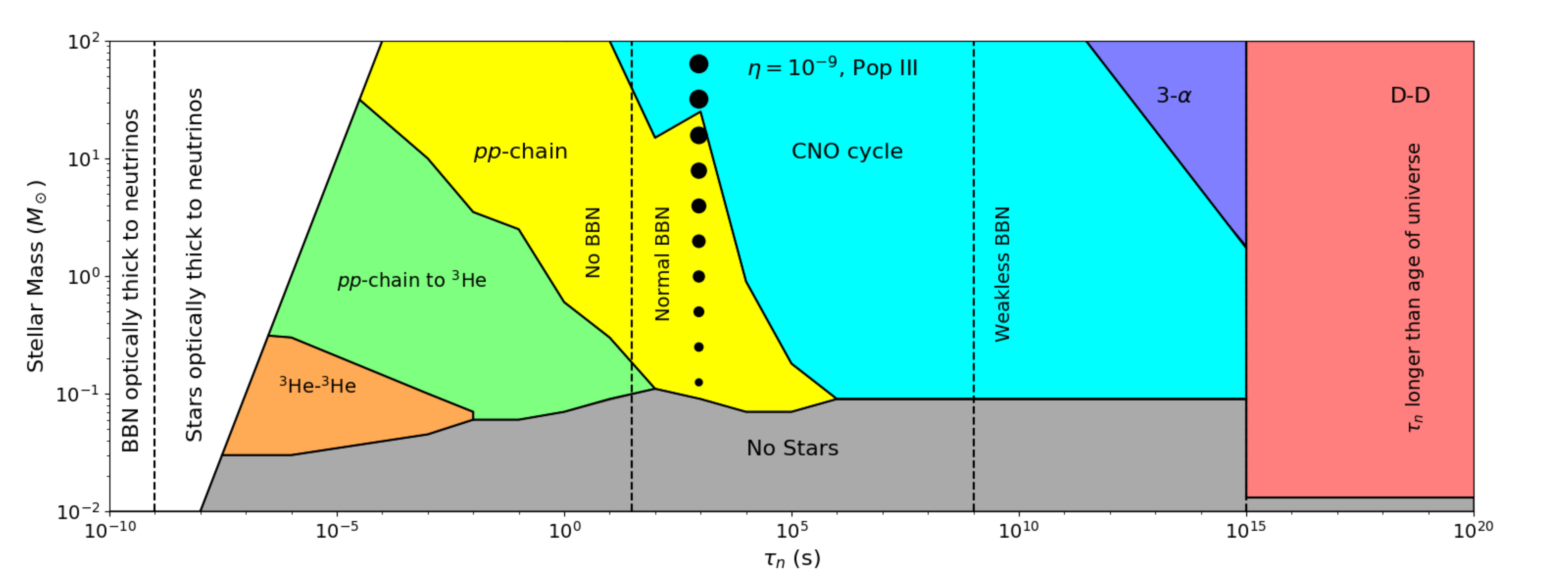}
	\caption{Regions denoting dominant nuclear processes in Population III stars in the space of stellar mass and neutron lifetime. We define the dominant process as the process that powers the star for the longest fraction of its lifetime, so some adjacent regions may undergo the same series of stages of nuclear burning, but for different periods of time. For this plot, the baryon-to-photon ratio, $\eta$ of our universe is maintained, such that regions with high neutron-lifetimes are extremely $^4$He-rich. The gray region denotes parts of the parameter space where long-lived nuclear-powered stars cannot exist. In the white region, the optical depth of stars (and BBN at the far left) to neutrinos is greater than unity, and we cannot simulate nuclear processes with the new neutrino interactions that this introduces.}
	\label{starmapsZ0}
\end{figure*}

\begin{figure*}[ht]
	\includegraphics[width=0.99\textwidth]{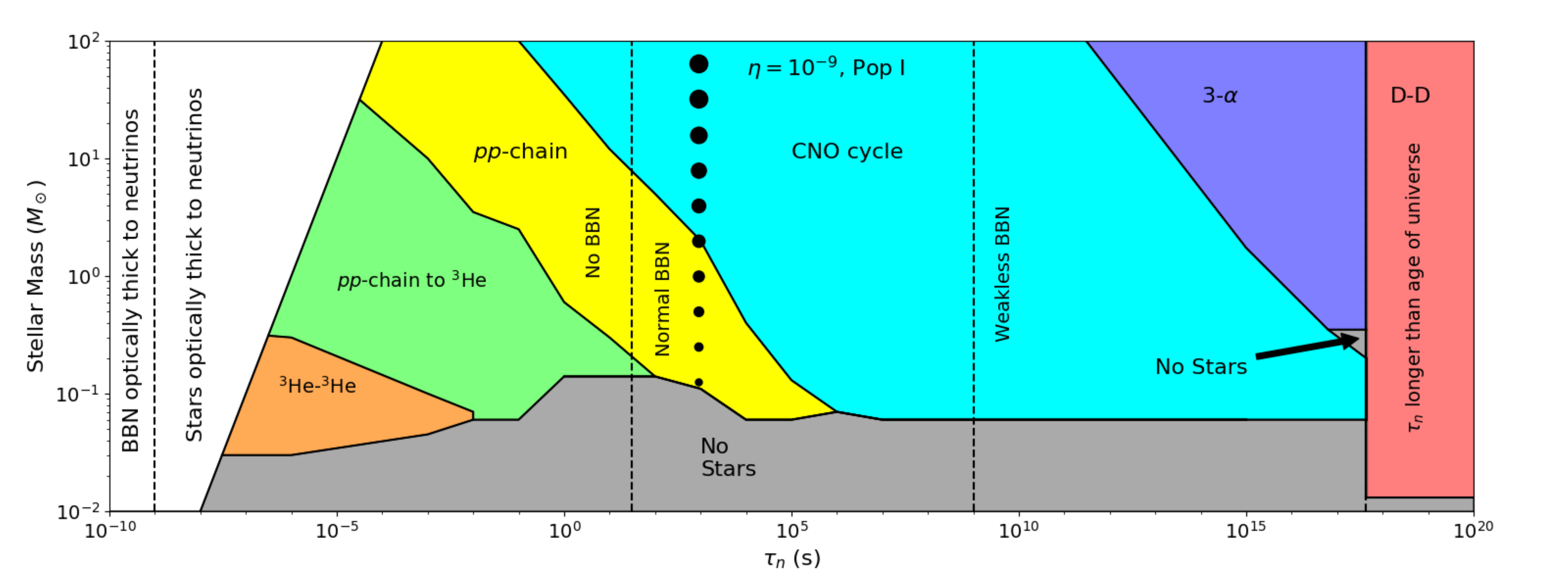}
	\caption{Same as Figure \ref{starmapsZ0}, but for Population I stars, resulting in a more prominent CNO cycle and more neutron decay by the time of star formation.}
	\label{starmapsZ2}
\end{figure*}

\begin{figure*}[ht]
	\includegraphics[width=0.99\textwidth]{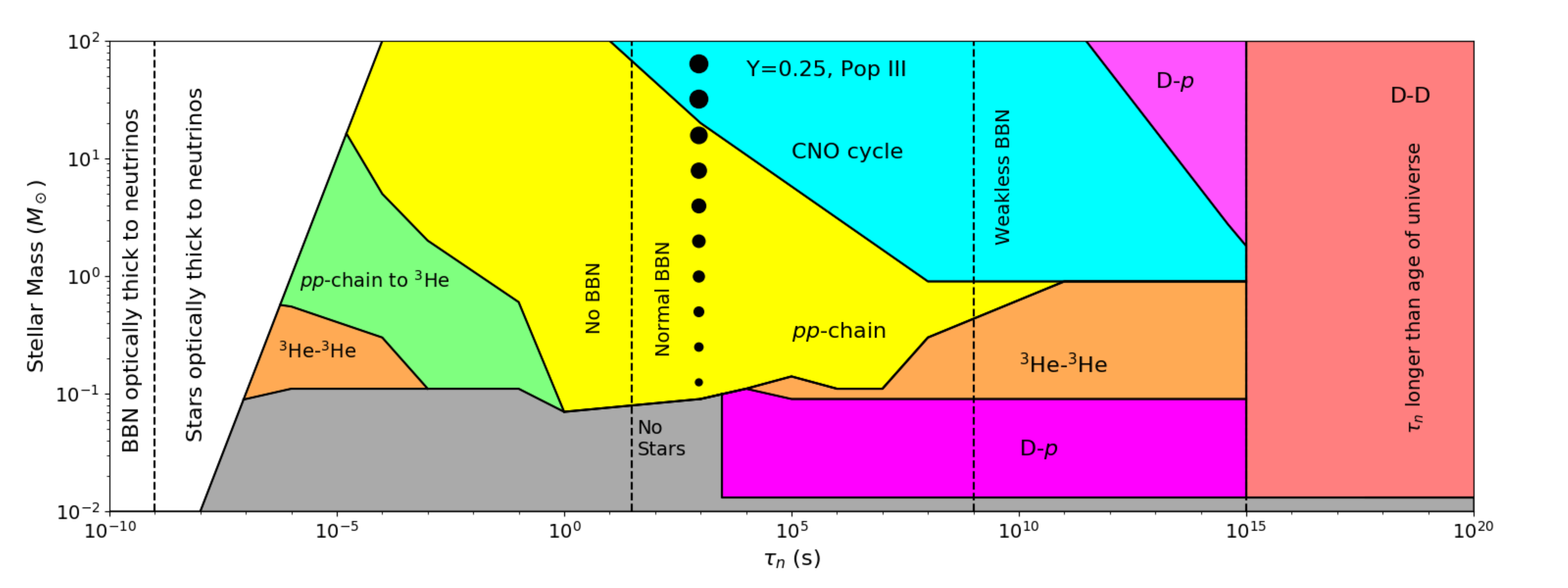}
	\caption{Same as Figure \ref{starmapsZ0}, but with $\eta$ allowed to vary so that the helium fraction from BBN, $Y_P$, is held constant. In this scenario, if $\tau_n>3000$ s, brown dwarfs will form with enough deuterium to become long-lived nuclear-burning objects, defined as stars for the purposes of this paper.}
	\label{starmapsX75}
\end{figure*}

\begin{figure*}[ht]
	\includegraphics[width=0.99\textwidth]{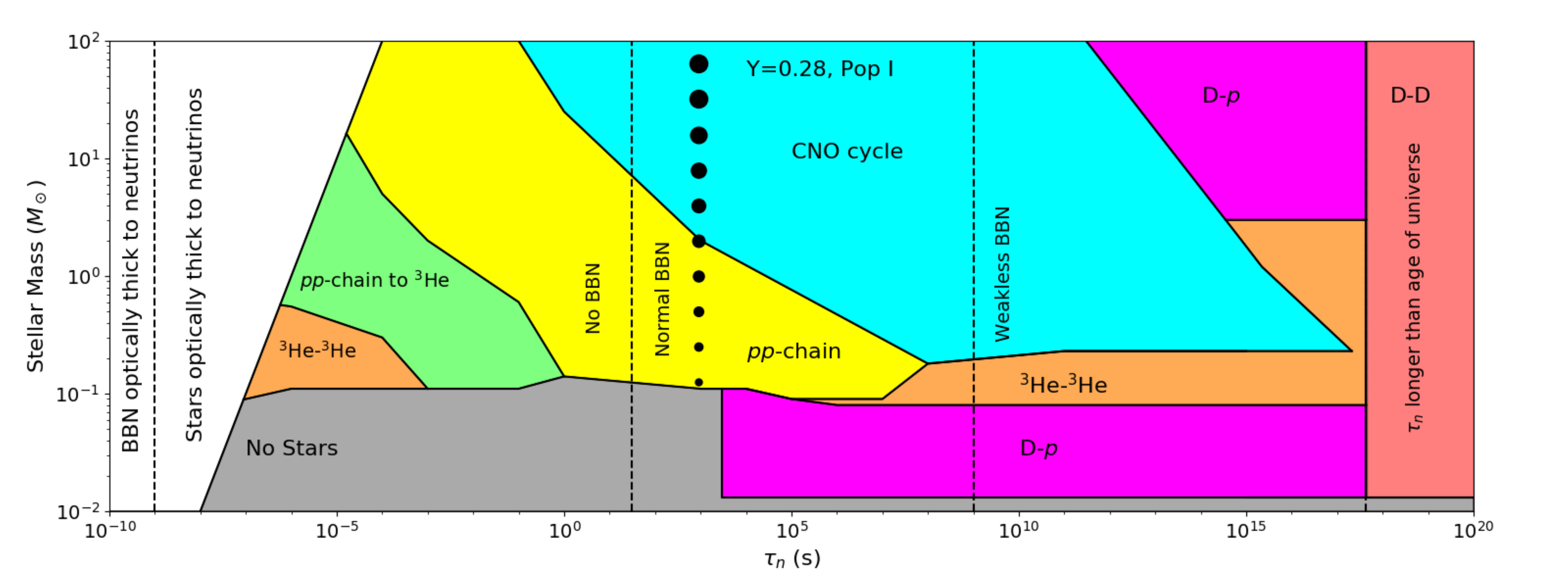}
	\caption{Same as Figure \ref{starmapsX75}, but for Population I stars, resulting in a more prominent CNO cycle and more neutron decay by the time of star formation.}
	\label{starmapsX70}
\end{figure*}

Stars can be classified into distinct domains in $M-\tau_n$-space according to the most important nuclear process that power them. Perhaps surprisingly, a large part of the parameter space allows long-lived stars to form, although the nuclear processes may vary greatly. We map out these domains in all four of our cases in Figures \ref{starmapsZ0}-\ref{starmapsX70}.

A few of the domains are nearly identical in all four cases. If the neutron lifetime is shorter than in our universe, the $pp$-chain will be more rapid, and main sequence stars will run more efficiently, although nuclear burning rates will not be significantly faster. Stars will reach an equilibrium sooner by switching on earlier in the collapse process, resulting in slightly cooler, but brighter stars, analogous to strong-burning stars. The effect on the minimum stellar mass appears to be statistically insignificant and obscured by the limitations of \mesa in calculating models for stars so far from those in our universe. However, one notable effect is that with the $pp$-chain being more efficient, it dominates over the CNO cycle at much higher masses. For Population I stars, the $pp$-chain dominates up to $\sim$20 $M_\odot$.

We note that in \mesa there can be a considerable margin of error in the boundaries of the domains, most notably in the minimum stellar mass. For example, for stars in our universe, the smallest mass that \mesa correctly simulates with hydrogen fusion is 0.12 $M_\odot$, compared with 0.08 $M_\odot$ for real stars. Thus, the jagged edges seen in a few of the boundaries are probably artifacts of the code and not statistically significant.

Another effect of a stronger weak force is that if $\tau_n<30$ s, then BBN will process nearly all baryons into free protons because neutrons will decay before nuclei can form. However, this will only have the effect of causing the first generation of stars to be made of 100\% hydrogen. (For this range, we must set $Y_P=0$ in all four cases.)

For even shorter neutron lifetimes, the $p$($p$,$e^+$)D step of the $pp$-chain will be faster than the $^3$He($^3$He,$2p$)$^4$He step, and hydrogen burning will occur in two stages: first the $pp$ and D-$p$ reactions to produce $^3$He, and second, $^3$He-burning to produce $^4$He. The point at which this occurs varies with mass from $\tau_n\sim1$ s for the smallest stars to $\tau_n\sim10^{-6}$ s for the largest, as the reaction rates vary differently with temperature.

Interestingly, the $^3$He-burning stage is fairly consistent in timescale regardless of $\tau_n$; for a 1 $M_\odot$ star, it consistently lasts about 500 Myr. Thus, in some parts of the parameter space (specifically low mass and a short neutron lifetime), the $^3$He-burning stage of a star's lifecycle will last longer, while in other parts, the $pp$-burning stage will last longer. Because we are addressing the phenomenological aspect of how stars appear in a weakful universe for the largest portion of their lives, we color these two cases as separate regions.

At the opposite end of the scale, if the neutron lifetime is longer than the age of the universe (for which we have adopted the somewhat short $10^{15}$ seconds for Population III stars and the present age of the universe for Population I stars), stars will form with a majority of their hydrogen in the form of deuterium and will be powered by D-D fusion. Such universes will approach the appearance of the weakless universe discuss in Ref. \cite{paperI}.

The most important difference between the Population III ($Z=0$) and Population I ($Z=0.02$) cases is that the CNO cycle is more dominant in Population I stars. The CNO cycle can still operate in metal-free stars, however. This occurs if the $pp$-chain is weak enough that the star's core reaches a temperature of 100 MK before $pp$-burning begins. At this temperature, the 3-$\alpha$ process rate will be sufficient to produce a trace ($\sim10^{-9}$) of CNO material. This 3-$\alpha$ rate is not sufficient to support the star on its own, but at this temperature, the CNO rate is high enough to do so even for trace amounts \cite{AdamsDeuterium}. The other major difference in the young universe where Population III stars form, stars can form with large amounts of deuterium with shorter neutron lifetimes (and thus fewer neutrons to form deuterium) than Population I stars.

There are two major differences between the constant-$\eta$ and constant-$Y_P$ cases. First in the constant-$\eta$ case, the deuterium fraction produced by BBN never exceeds 2\%, so we do not count any low-mass deuterium-burning stars in these cases. The other difference is that over much of the parameter space, constant-$\eta$ stars are denser than in our universe. This is because stars in universes with a longer neutron lifetime and the same baryon abundance will form with a greater amount of helium. This higher density results in smaller minimum masses for all nuclear processes including, for example, allowing even low-mass stars to undergo CNO burning.

Generally, as the weak force is made weaker, stars will transition from the $pp$-chain to the CNO cycle as their main power source, with the crossover point (which is 1.3 $M_\odot$ in our universe) occurring at lower masses. At the same time, the minimum mass for $pp$-burning increases until it reaches the minimum mass for CNO burning. The minimum CNO-burning mass is lower than in our universe even for a similar composition (for which it falls at 0.23 $M_\odot$) because with a weaker $pp$-chain, the core is able to contract further and reach a higher temperature before reaching an equilibrium state. In the constant-$\eta$ cases, the minimum CNO-burning mass for the denser stars that occur there is even lower, at 0.06-0.09 $M_\odot$.

The rate of the CNO cycle normally depends on the strong-burning rates of proton capture reactions. However, it also involves two weak decays that are very fast on stellar timescales: $^{13}$N and $^{15}$O, both of which have lifetimes of minutes in our universe. As long as these decays remain fast relative to the lifetime of the star, the CNO cycle is not affected by the strength of the weak force. However, if these decays are longer than the CNO timescale, then the CNO cycle will be suppressed.

At solar metallicities, stars have on the order of 1 CNO nucleus for every $\sim$100 $^4$He nuclei produced. In stars where the CNO cycle dominates, this necessarily means that the CNO timescale is on the order of 1\% of the main sequence lifetime. For the most massive stars, the CNO timescale is thus on the order of a ten thousand years. The $^{13}$N and $^{15}$O decay times will be comparable to the CNO timescale for massive stars at $\tau_n\sim3\times10^{11}$ s, and longer for less massive stars. If the weak force is weaker than this, the CNO cycle shuts off, and the $pp$-chain is very weak, leaving the 3-$\alpha$ process as the dominant energy source for massive stars in this region of the parameter space.

For Population III stars, the CNO abundance will be much smaller at $\sim10^{-9}$, and the CNO timescale will be proportionately shorter: as little as $3\times10^4$ s for the most massive stars. The 3-$\alpha$ process will dominate at significantly shorter $\tau_n$ in this case.

The minimum mass for helium burning also depends on the stellar density. For the constant-$\eta$ case, in the helium-rich high-$\tau_n$ region where it is most important, the minimum mass is 0.35 $M_\odot$, similar to our universe. For the constant-$Y_P$ case, which is more hydrogen-rich, the minimum mass is 1.2 $M_\odot$. If the neutron lifetime is sufficiently long, stars smaller than this mass will form with the CNO cycle switched off.

These combined factors allow a potential gap in the stellar mass function for certain neutron lifetimes, where no significant nuclear processes are available to them. These are stars for which the neutron lifetime is too long for the CNO cycle to function, but too short for significant deuterium to remain to produce deuterium-powered stars, and are too small to fuse helium. This gap appears in the Population I region plot for the constant-$\eta$ case (Figure \ref{starmapsZ2}).

However, this applies differently to the constant-$Y_P$ case. Here, for $\tau_n>3000$ s, stars form with significant amounts of deuterium, approaching 10\%, and deuterium fusion becomes an important stellar process. Brown dwarf-mass objects will be low-mass D-$p$ burning stars by our definition. More massive stars larger than $\sim$0.1 $M_\odot$ will undergo a second stage of fusion, burning $^3$He to $^4$He. This process produces surplus protons, which can only be incorporated into $^4$He by being converted to neutrons by weak reactions, so still more massive stars will undergo a third CNO-burning stage, converting the remaining hydrogen to helium. This mass limit varies from 0.23 to 0.9 $M_\odot$ depending on metallicity. For these objects the CNO-burning phase is the longest and produces the most energy. In the upper right region were the 3-$\alpha$ process dominates for deuterium-poor stars, $^3$He-burning is again the dominant process for these deuterium-rich stars.

\subsection{Evolutionary Tracks}

\begin{figure}[ht]
	\includegraphics[width=0.49\textwidth]{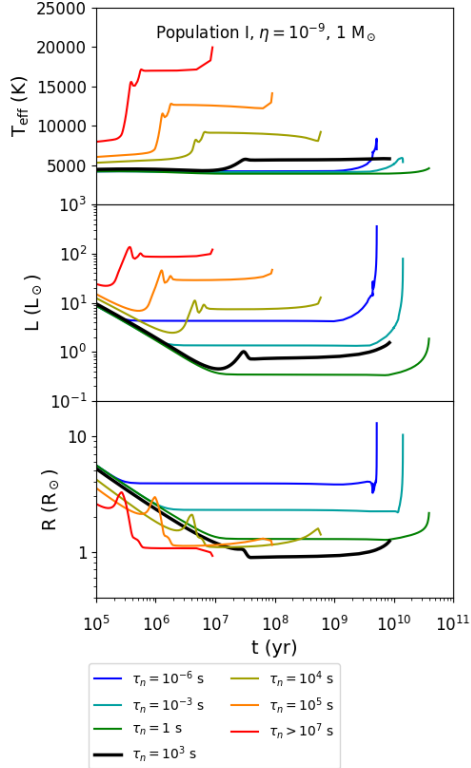}
	\caption{Evolution of the effective temperature, luminosity, and radius of a 1 $M_\odot$ star in universes with constant $\eta$ and a neutron lifetime ranging from 10$^{-6}$ s to 10$^7$ s. Neutron lifetimes longer than 10$^7$ s do not have a significant influence on stellar evolution. Most of the differences in stellar evolution between models are due to the helium fraction of the star.}
	\label{timeZ2}
\end{figure}

\begin{figure}[ht]
	\includegraphics[width=0.49\textwidth]{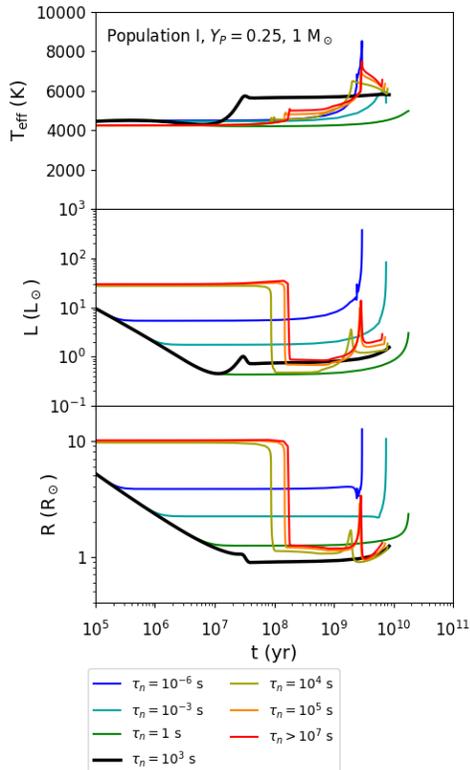}
	\caption{Same as Figure \ref{timeZ2}, but for universes with constant $Y_P$. Two distinct types of evolutionary tracks are seen: those for stars with $\tau_n>3000$ s, which form with significant amounts of deuterium, and those for stars with $\tau_n<3000$ s, which do not.}
	\label{timeX70}
\end{figure}

We examine the evolutionary tracks of stars in universes with a range of neutron lifetimes in Figures \ref{timeZ2} and \ref{timeX70}. These figures show the evolution of a 1 $M_\odot$ star in the constant-$\eta$ case and the constant-$Y_P$ case, respectively, both at solar metallicity. The zero-metallicity cases do not look dramatically different, except for being somewhat hotter and less luminous, as expected for zero-metallicity stars.

In the constant-$\eta$ case, the most significant change when varying the neutron lifetime is in the stellar lifetime. As described above, a longer neutron lifetime leads to a lower hydrogen fraction, which shortens the stellar lifetime. These low-hydrogen stars are also denser, causing them to be hotter and more compact, which further shortens the stellar lifetime. The main sequences approach a limit at $\tau_n > 10^7$ s, where a 1 $M_\odot$ star has a lifetime of $\sim$10 Myr, a luminosity of $\sim$100 $L_\odot$, and a surface temperature of $\sim$17,000 K. Longer neutron lifetimes result in main sequences nearly identical to $\tau_n \sim 10^7$ s.

As the neutron lifetime is decreased and hydrogen fusion becomes more efficient along with hydrogen becoming more abundant, stars are longer lived, cooler, and fainter. In particular, the luminosity reaches a minimum at $\tau_n$ = 1-100 s, where the initial hydrogen fraction is 100\%. Beyond this point, as the ignition temperature of hydrogen becomes still lower, fusion begins and the stars reach equilibrium earlier in the collapse process, resulting in stars that are larger, brighter, and cooler than in our universe, approaching the properties of red giant-like strong-burning stars like those in a weakless universe.

In the constant-$Y_P$ case on the other hand, stellar lifetime does not vary greatly with $\tau_n$, all falling in the range of 1.5-20 Gyr, because the initial composition is not radiacally different except for the deuterium abundance. For universes more weakful than our own, we see the same progression of larger, redder stars as the neutron lifetime decreases. However, for less weakful universes, we see a single, distinctive evolutionary track for all of the models we plot. At constant $Y_P$, 1 $M_\odot$ stars have very similar compositions and the same nuclear processes operating for $\tau_n$ ranging from $10^4$ s to $10^{15}$ s. These stars show the three phases of nuclear burning described above.

For these high-deuterium stars, deuterium burning ignites early in the contraction process, resulting in a large, cool, red giant-like star for the first $\sim$100 Myr of its life. Once the deuterium is exhausted, the star rapidly contracts to a near-main sequence state in which $^3$He-burning occurs, lasting 2-3 Gyr. (This is the most analogous stage to the main sequence in our universe.) This is followed by a miniature giant-like expansion phase in which the star brightens by a factor of $\sim$20 before core hydrogen burning begins (in this case, the CNO cycle), and the star falls back to the main sequence for the longest stage of its life cycle, lasting 4-5 Gyr. After this, post-main-sequence evolution occurs similarly to our universe, since this stage is powered by strong-burning reactions.

\begin{figure}[htb]
	\includegraphics[width=0.99\columnwidth]{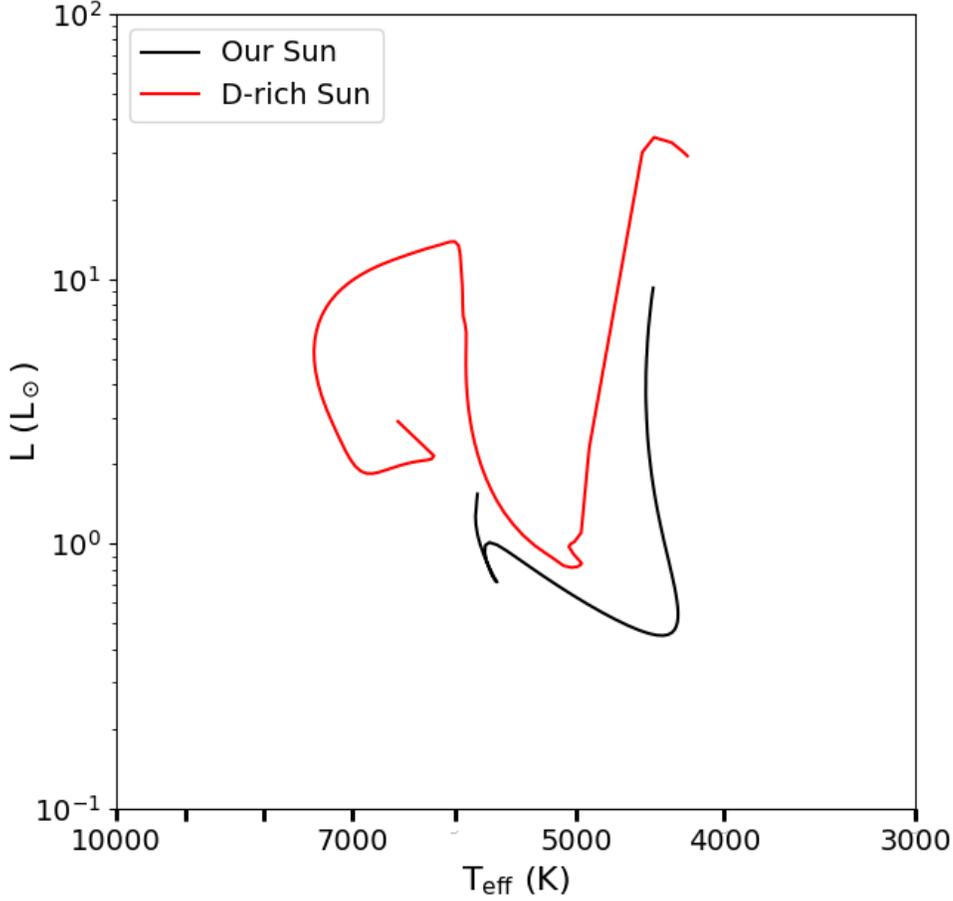}
	\caption{The two general types of evolutionary tracks for a 1 $M_\odot$ star in universes with constant $Y_P$ plotted on the H-R diagram. The black curve is the evolutionary track for our Sun, more generally representing stars with $\tau_n<3000$ s. The red curve is the evolutionary track for $\tau_n=10^8$ s, and more generally represents stars with $\tau_n>3000$ s.}
	\label{evoplot}
\end{figure}

To give a clearer picture of the evolution of these deuterium-rich stars, in Figure \ref{evoplot}, we plot the evolutionary track for a representative 1 $M_\odot$ deuterium-rich star on the H-R diagram compared with the evolutionary track for our Sun up to the end of the main sequence. The deuterium-rich star descends the Hayashi track only after the end of deuterium burning to land on the main sequences during $^3$He burning. The star then expands along a small giant-like branch before contracting again to land higher on the main sequence during core hydrogen burning.

% % % % % % % % % % % % % % % % % % % % % % % % % % % % % % % % % % % % % % % % % % % % % % % % % % % % % % % % % % % % % % % % % % %

\section{Chemical Evolution and Habitability}

For most the parameter space considered in this study, chemical
evolution proceeds similarly to that in our universe. The same nuclear
reactions provide power in stellar cores, especially during post-main
sequence evolution, although they proceed at different rates. As a
result, only the abundances of the lightest isotopes are significantly
affected by the strength of the weak force. A stronger weak force
results in a lower helium fraction due to much lower helium production
during BBN. Such universes could also have a dramatically
higher cosmic $^3$He abundance because in a large region of the
parameter space, $^3$He production occurs separately from $^4$He
production.
Meanwhile, a weaker weak force, on the other hand, can result in a
lower light hydrogen (proton) abundance and a higher deuterium
abundance from BBN.

A stronger weak force would result in core-collapse supernovae being more efficient and dispersing more oxygen and other $\alpha$ elements because the neutrinos that power them would interact more strongly. However, this trend may reverse for an even stronger weak force as neutrinos would become trapped in the core and thus unable to impart sufficient momentum to the overlying gas layers to eject them, which would then cause the supernova to fail \cite{carrrees}. If core-collapse supernovae do function, a stronger weak force would also cause $r$-process yields from neutron star mergers to decrease as \bdecay becomes faster, and the $r$-process comes to more closely resemble the $s$-process. Since the $r$-process timescale is $\sim$100 ms \cite{Surman13}, and typical \bdecay times on the neutron-rich side of the line of stability are on the order of 10 s to $10^4$ s, the $r$-process will closely resemble the $s$-process at $\tau_n \lesssim 10^{-2}$ s.

For a sufficiently weak weak force, one important change occurs to chemical evolution in the CNO cycle. If the lifetimes of $^{13}$N and $^{15}$O are similar to the CNO timescale, it will both slow the CNO cycle as described above, and also open up new reaction pathways in the various branches of the cycle:
\begin{align}
^{13}\rm{N} + p &\rightarrow\,^{14}\rm{O} \\ \nonumber
^{14}\rm{O} &\rightarrow\,^{14}\rm{N} + e^+ + \nu_e \\ \nonumber
^{17}\rm{F} + p &\rightarrow\,^{18}\rm{Ne} \\ \nonumber
^{18}\rm{Ne} &\rightarrow\,^{16}\rm{O} + 2p \\ \nonumber
^{18}\rm{F} + p &\rightarrow\,^{19}\rm{Ne} \\ \nonumber
^{19}\rm{Ne}&\rightarrow\,^{19}\rm{F} + e^+ + \nu_e,
\end{align}
and so on. With more possibilities to continue adding more protons, it is unclear where the CNO cycle will halt under typical stellar conditions by ejecting an alpha particle as in the $^{15}\rm{N} + p\rightarrow\, ^{12}\rm{C} +\,^4\rm{He}$ step of the CNO-I branch. It may halt at fluorine, as in the hot CNO cycle in our universe, or it may continue further under cooler conditions. It is plausible that this would result in significantly higher fluorine and neon abundances and less carbon than in our universe.

Another effect of a weaker weak force is that weaker neutrino interactions are likely to cause core-collapse supernovae to fail, as in the weakless universe. In the weakless universe, elements heavier than oxygen are dispersed essentially only by Type Ia supernovae. The core-collapse supernovae needed to produce the $\alpha$-elements and the neutron stars needed to enact the $r$-process will not occur in a significantly weaker universe. However, with the weak force still included, the $s$-process would remain in operation, allowing an additional pathway for heavy elements, although its yields may be lower. The $s$-process will also come to resemble the $r$-process if the neutron capture timescale is shorter than typical \bdecay times. The $s$-process timescle is 5-100 years \cite{Matsuda80}, so this will occur for $\tau_n \gtrsim 10^{12}$ s.

Another important consideration to habitability in a universe with a different weak force is stellar lifetime. In the majority of the parameter space, long-lived stars can exist with similar chemical evolution to our universe, and this presents no significant barrier to habitability. However, in a large minority of the parameter space with large $\eta$ and large $\tau_n$, stars are not long-lived enough for life as we know it to develop due to their low hydrogen fractions. As in the weakless universe, a low value of $\eta$ is needed to allow a large hydrogen abundance to survive BBN. However, in this case, with the weak force still operating, stars in these universes are more hostile to life with a long, early deuterium-burning stage with high luminosity, along with the period that would be the main sequence lifetime of stars in our universe interrupted by an expansion phase between $^3$He burning and core hydrogen burning. With such variability in stellar brightness, it is not clear whether a planet orbiting such a star could acquire and retain a large volatile reservoir long enough for life as we know it to develop. It is possible, then, that life in such a universe would be relegated to low-mass stars with deuterium-burning times longer than 1 Gyr that can provide a relatively stable environment, as in the weakless universe.

% % % % % % % % % % % % % % % % % % % % % % % % % % % % % % % % % % % % % % % % % % % % % % % % % % % % % % % % % % % % % % % % % % %

\section{Conclusion}
\label{sec:conclude}

This paper has considered a class of universes for which the strength of the weak force is vastly different from that in our own region of space-time. We have performed numerical simulations for both Big Bang Nucleosynthesis and stellar evolution using a wide range of possible strengths for the weak force. The overall finding of this study is that universes are remarkably robust to changes in the strength of the weak force, in that a wide range of universes remain viable. Here we provide a summary of our specific results (Section \ref{sec:summary}) along with a discussion of their implications (Section \ref{sec:discuss}).

\subsection{Summary of Results} 
\label{sec:summary} 

For the sake of definiteness, we parameterize the strength of the weak force in terms of the neutron lifetime (where $\tau_n \simeq 885$ sec for our universe). The limit where $\tau_n\to\infty$ corresponds to the weakless universe, as would a neutron lifetime longer than the age of the universe, $\tau_n>10^{18}$ s. At the opposite end of the scale, $\tau_n\le10^{-9}$ sec leads to substantial changes to nuclear structure. Overall, we consider a range of $\tau_n$ spanning more than 25 orders of magnitude.

We have studied the epoch of Big Bang Nucleosynthesis in detail, using the code \burst. In the limit of large $\tau_n$, neutrons do not decay during the epoch of BBN and the universe tends to produce large amount of helium. In order to allow some protons to survive, which is necessary for the universe to have nuclear fuel and produce water, the baryon to photon ratio must be smaller than that of our universe. For small $\eta$ and large $\tau_n$, BBN allows for substantial amounts of protons, deuterium, and helium. In the limit of small $\tau_n\lesssim10$ sec, neutrons decay so quickly that the universe emerges from the BBN epoch with an almost pure hydrogen composition.

We have studied stellar structure and evolution using the state of the art code \mesa. Here, the parameter space for stars is broken into four regimes characterized by two variables. The first variable is metallicity, for which we consider both $Z=0$ (Population III) and $Z=0.02$ (Population I) metallicities. The second variable determines the starting stellar composition. Here we consider the abundances produced by BBN for constant $\eta$, as well as composition characterized by constant helium mass fraction.

Over the allowed range of strengths of the weak force, equivalently varying values of $\tau_n$, energy generation in stellar interiors is dominated by a variety of different nuclear reactions (see Figures 2$-$5). In the limit of large $\tau_n$, universes tend to produce large abundances of deuterium, and stellar energy generation is dominated by deuterium burning. As the weak force becomes stronger, so that $\tau_n$ is shorter, the $pp$-chain becomes more effective than the CNO cycle and comes to dominate even at high stellar masses. For sufficiently small values of $\tau_n$, the $pp$-reaction is fast enough (relative to subsequent reactions in the standard $pp$-chain) that stars experience a new nuclear burning phase that produces helium-3 as its end product (which is only later burned into helium-4). The boundaries between the different regimes of nuclear burning depend on stellar mass, as expected. 

The allowed range of stellar masses is largely insensitive to the strength of the weak force. The upper mass limit, taken here to be $M_{\rm max}=100M_\odot$, is set by radiation pressure and other considerations, and does not depend on $\tau_n$. The lower mass limit (near 0.1 $M_\odot$) is nearly constant for the case of fixed $\eta$. For constant helium abundance, universes with larger $\tau_n$ produce substantial deuterium abundances, which allow small stars to operate through deuterium fusion (D-$p$ burning). As result, a window opens up for small stars with masses 0.01$-$0.1 $M_\odot$ in universes with $\tau_n\gtrsim3000$ sec. These deuterium-rich stars become larger, brighter, and redder at all masses, and thus resemble red giants in our universe. For universes with constant $\eta$, a large fraction of the hydrogen fuel is already processed into helium during the epoch of BBN, so that stars are hotter, brighter, and shorter lived.

In the opposite limit with small $\tau_n$, stars operate by burning protons into helium-3 at moderately lower temperatures. Their radii and luminosities are comparable to, but somewhat larger than, those of stars in our universe, whereas their surface temperatures are relatively unchanged. Consistent with this insensitivity of stellar properties, the main sequences at constant $\eta$ in the H-R diagram are much like those observed in our universe. Stellar lifetimes in such universes will be modestly shorter than in our universe, by as much as a factor of a few, although this will be partially mitigated by the higher $^1$H content.

\subsection{Discussion} 
\label{sec:discuss} 

The results of this paper show that universes can remain potentially habitable over a wide range of strengths for the weak force. Given the resilient nature of universes subject to these variations, it is useful to consider the range of properties that could render the universe sterile.

In Fig.\ \ref{fig:neff_vs_mntau}, we noted that our universe exists in the
middle of a plateau where $\neff\simeq3$.  The neutron lifetime must change by roughly two
orders of magnitude in either direction for \neff to deviate from 3.0.  The
plateau is a result from the fact that neutrino decoupling occurs before
$e^\pm$-annihilation and after $\mu^\pm$-annihilation.  As we always assume the
charged lepton seas are in thermal and chemical equilibrium with the plasma,
the mass of the respective particles defines the relevant energy scale for the two
epochs.  This location of our universe on the plateau is partly apparent from
the relation that $(G_F^2\mpl)^{-1/3}\sim1\,{\rm MeV}$, an energy scale between
the two lepton mass scales (this is similar to the decoupling energy scale
which we took to be $3.0\,{\rm MeV}$ in Sec.\ \ref{ssec:model}). For our universe, 
interesting consequences could arise if there exist additional interactions that affect 
the temperature scale where neutrinos begin to decouple. 
Neutrino magnetic moments \cite{Vassh_tdecoup} or hidden
interactions \cite{2005JHEP...12..021B} could lower the decoupling temperature.
Figure \ref{fig:neff_vs_mntau} shows that such an interaction would have to
overwhelm the weak interaction by two orders of magnitude to produce changes in
\neff, although secret interactions solely within the neutrino seas could have
other ramifications on the Cosmic Microwave Background and large scale structure
\cite{2016JCAP...01..007B,2017JCAP...07..038F}.  A changing \neff would change
the epoch of matter-radiation equality, which would have ramifications for the
onset of and galaxy formation \cite{2018MNRAS.477.3727B} and large scale structure 
\cite{AAGM:2017_vac}. However, neither the low nor high bounds of \neff
in Eqs.\ \eqref{eq:neff_no_mu} and \eqref{eq:neff_w_e} will dramatically shift
the matter-radiation equality before habitability becomes an issue.  In light
of these facts, we conclude that there does not exist a fine-tuning argument
for how $G_F$ affects the expansion history of the universe.  What is
more sensitive to the weak interaction is the primordial abundances,
specifically the ratio of hydrogen to helium.  An increase in \taun of two
orders-of-magnitude will flip the universe from hydrogen dominated to helium
dominated.  We would presume that life requires a nonzero primordial hydrogen
component, although we do not know what the strict lower bound may be.
Nevertheless, if \taun is large enough a dearth of hydrogen could be
problematic for life.

One way to inhibit life is for the universe to emerge from BBN with few unbound protons. In the limit of $\tau_n\to\infty$, this occurs at and even somewhat below the value $\eta_0$ in our universe. Specifically, for $\eta=\eta_0$ = $6\times10^{-10}$, only $\sim1\%$ of the mass of the universe remains in protons \cite{paperI}. However, universes with $\eta<\eta_0$ remain viable even for $\tau_n\to\infty$. We find that it is relatively difficult for BBN processes to compromise the universe, consistent with previous results \cite{AAGM:2017_vac}.

The opposite limit of stronger weak forces is more problematic. Stars cease to operate normally if the neutrinos produced by nuclear reactions become optically thick. This limit depends on stellar mass, but requires $\tau_n\gtrsim10^{-6}$ sec for solar type stars. Even smaller values of $\tau_n$, corresponding to an even stronger weak force, lead to substantial contributions to the energy budget of nuclei. The strength of the weak interaction required to compromise nuclear structure remains unknown, but the periodic table is likely to be quite different for the regime where $\tau_n\ll10^{-10}$ sec. In this regime, neutrinos are also optically thick during BBN, so the output of BBN is equally uncertain.

As outlined above, universes continue to be viable when the strength of the weak force is varied over many orders of magnitude. In addition to its ramifications for fine-tuning, these results also provide us with a deeper understanding of how BBN and stellar evolution operate in our universe. One general trend emerging from this study is that stars are more robust than nuclei. If the laws of physics$-$in this context the strength of the weak force$-$allow for complex nuclei to exist, then stellar interiors are likely to produce them. The reason for this flexibility is that stars can operate using a wide range of different nuclear processes, from the $pp$-reaction through the weak interaction to deuterium burning through the strong interaction (and many additional chains in between). Moreover, stars span a wide range of mass (a factor of $\sim1000$) and can produce an even wider range of densities and temperatures in their cores. This enormous range of available parameter space, coupled with the exponential temperature sensitivity of nuclear reaction rates, allows for stars to operate over a wide range of realizations of the laws of physics (see also \cite{AdamsDeuterium,AdamsStars}). 

%As long as stars get to light hydrogen burning, you pretty much get the main sequence.

% % % % % % % % % % % % % % % % % % % % % % % % % % % % % % % % % % % % % % % % % % % % % % % % % % % % % % % % % % % % % % % % % % %

\acknowledgements
We would like to thank Daniele Alves, Vincenzo Cirigliano, George Fuller and Lillian Huang for useful discussions. This work was supported by the John Templeton foundation through Grant ID55112: Astrophysical Structures in other Universes, and by the University of Michigan. This work was supported in part by the Los Alamos National Laboratory Institutional Computing Program, under U.S. Department of Energy National Nuclear Security Administration Award No. DE-AC52-06NA25396. Computational resources and services were also provided by Advanced Research Computing at the University of Michigan. EG acknowledges support from the National Science Foundation, Grant PHY-1630782, and the Heising-Simons Foundation, Grant 2017-228.

% % % % % % % % % % % % % % % % % % % % % % % % % % % % % % % % % % % % % % % % % % % % % % % % % % % % % % % % % % % % % % % % % % %

%\clearpage
\bibliographystyle{apj}
%\bibliography{apj-jour,refs}

\begin{thebibliography}{48}
	\expandafter\ifx\csname natexlab\endcsname\relax\def\natexlab#1{#1}\fi
	
	\bibitem[{{Adams}(2016)}]{AdamsStars}
	{Adams}, F.~C. 2016, \jcap, 2, 042
	
	\bibitem[{{Adams} {et~al.}(2017){Adams}, {Alexander}, {Grohs}, \&
		{Mersini-Houghton}}]{AAGM:2017_vac}
	{Adams}, F.~C., {Alexander}, S., {Grohs}, E., \& {Mersini-Houghton}, L. 2017,
	\jcap, 3, 021
	
	\bibitem[{{Adams} \& {Grohs}(2017)}]{AdamsDeuterium}
	{Adams}, F.~C., \& {Grohs}, E. 2017, Astroparticle Physics, 91, 90
	
	\bibitem[{{Agrawal} {et~al.}(1998){Agrawal}, {Barr}, {Donoghue}, \&
		{Seckel}}]{1998PhRvD..57.5480A}
	{Agrawal}, V., {Barr}, S.~M., {Donoghue}, J.~F., \& {Seckel}, D. 1998, \prd,
	57, 5480
	
	\bibitem[{{Akhmedov} {et~al.}(1998){Akhmedov}, {Rubakov}, \&
		{Smirnov}}]{1998PhRvL..81.1359A}
	{Akhmedov}, E.~K., {Rubakov}, V.~A., \& {Smirnov}, A.~Y. 1998, Physical Review
	Letters, 81, 1359
	
	\bibitem[{{Bahcall}(1995)}]{Bahcall95}
	{Bahcall}, J.~N. 1995, in Solar Modeling, ed. A.~B. {Balantekin} \& J.~N.
	{Bahcall}, 1
	
	\bibitem[{{Barnes} {et~al.}(2018){Barnes}, {Elahi}, {Salcido}, {Bower},
		{Lewis}, {Theuns}, {Schaller}, {Crain}, \& {Schaye}}]{2018MNRAS.477.3727B}
	{Barnes}, L.~A., {et~al.} 2018, \mnras, 477, 3727
	
	\bibitem[{{Barranco} {et~al.}(2005){Barranco}, {Miranda}, \&
		{Rashba}}]{2005JHEP...12..021B}
	{Barranco}, J., {Miranda}, O.~G., \& {Rashba}, T.~I. 2005, Journal of High
	Energy Physics, 12, 021
	
	\bibitem[{{Barrow} \& {Tipler}(1986)}]{bartip}
	{Barrow}, J.~D., \& {Tipler}, F.~J. 1986, {The anthropic cosmological
		principle}
	
	\bibitem[{{Baumann} {et~al.}(2016){Baumann}, {Green}, {Meyers}, \&
		{Wallisch}}]{2016JCAP...01..007B}
	{Baumann}, D., {Green}, D., {Meyers}, J., \& {Wallisch}, B. 2016, \jcap, 1, 007
	
	\bibitem[{{Bedaque} {et~al.}(2011){Bedaque}, {Luu}, \&
		{Platter}}]{2011PhRvC..83d5803B}
	{Bedaque}, P.~F., {Luu}, T., \& {Platter}, L. 2011, \prc, 83, 045803
	
	\bibitem[{{Bertulani} \& {Kajino}(2016)}]{2016PrPNP..89...56B}
	{Bertulani}, C.~A., \& {Kajino}, T. 2016, Progress in Particle and Nuclear
	Physics, 89, 56
	
	\bibitem[{{{\bf CMB-S4} Collaboration, J.~E. Carlstrom {\it et
				al}}.(2016)}]{cmbs4_science_book}
	{{\bf CMB-S4} Collaboration, J.~E. Carlstrom {\it et al}}. 2016, ArXiv e-prints
	
	\bibitem[{{Blaschke} \& {Cirigliano}(2016)}]{Blaschke_Cirigliano_2016}
	{Blaschke}, D.~N., \& {Cirigliano}, V. 2016, \prd, 94, 033009
	
	\bibitem[{{Carr} \& {Rees}(1979)}]{carrrees}
	{Carr}, B.~J., \& {Rees}, M.~J. 1979, \nat, 278, 605
	
	\bibitem[{{Carter}(1983)}]{carter}
	{Carter}, B. 1983, Philosophical Transactions of the Royal Society of London
	Series A, 310, 347
	
	\bibitem[{Davidson {et~al.}(2008)Davidson, Nardi, \& Nir}]{Davidson:2008bu}
	Davidson, S., Nardi, E., \& Nir, Y. 2008, Phys. Rept., 466, 105
	
	\bibitem[{{Davies}(2004)}]{davies}
	{Davies}, P.~C.~W. 2004, Modern Physics Letters A, 19, 727
	
	\bibitem[{{Dicus} {et~al.}(1982){Dicus}, {Kolb}, {Gleeson}, {Sudarshan},
		{Teplitz}, \& {Turner}}]{1982PhRvD..26.2694D}
	{Dicus}, D.~A., {Kolb}, E.~W., {Gleeson}, A.~M., {Sudarshan}, E.~C.~G.,
	{Teplitz}, V.~L., \& {Turner}, M.~S. 1982, \prd, 26, 2694
	
	\bibitem[{{Donoghue}(2016)}]{Donoghue16}
	{Donoghue}, J.~F. 2016, Annual Review of Nuclear and Particle Science, 66, 1
	
	\bibitem[{Drewes {et~al.}(2018)Drewes, Garbrecht, Hernandez, Kekic,
		Lopez-Pavon, Racker, Rius, Salvado, \& Teresi}]{Drewes:2017zyw}
	Drewes, M., {et~al.} 2018, Int. J. Mod. Phys., A33, 1842002
	
	\bibitem[{{Ellis} {et~al.}(2004){Ellis}, {Kirchner}, \& {Stoeger}}]{Ellis04}
	{Ellis}, G.~F.~R., {Kirchner}, U., \& {Stoeger}, W.~R. 2004, \mnras, 347, 921
	
	\bibitem[{{Forastieri} {et~al.}(2017){Forastieri}, {Lattanzi}, {Mangano},
		{Mirizzi}, {Natoli}, \& {Saviano}}]{2017JCAP...07..038F}
	{Forastieri}, F., {Lattanzi}, M., {Mangano}, G., {Mirizzi}, A., {Natoli}, P.,
	\& {Saviano}, N. 2017, Journal of Cosmology and Astro-Particle Physics, 2017,
	038
	
	\bibitem[{{Fukugita} \& {Yanagida}(1986)}]{1986PhLB..174...45F}
	{Fukugita}, M., \& {Yanagida}, T. 1986, Physics Letters B, 174, 45
	
	\bibitem[{{Fuller} {et~al.}(1980){Fuller}, {Fowler}, \& {Newman}}]{FFN:I}
	{Fuller}, G.~M., {Fowler}, W.~A., \& {Newman}, M.~J. 1980, \apjs, 42, 447
	
	\bibitem[{{Grohs} \& {Fuller}(2016)}]{WFO_approx}
	{Grohs}, E., \& {Fuller}, G.~M. 2016, Nuclear Physics B, 911, 955
	
	\bibitem[{{Grohs} \& {Fuller}(2017)}]{xmelec}
	---. 2017, Nuclear Physics B, 923, 222
	
	\bibitem[{{Grohs} {et~al.}(2015){Grohs}, {Fuller}, {Kishimoto}, \&
		{Paris}}]{GFKP-5pts:2014mn}
	{Grohs}, E., {Fuller}, G.~M., {Kishimoto}, C.~T., \& {Paris}, M.~W. 2015,
	\jcap, 5, 017
	
	\bibitem[{{Grohs} {et~al.}(2018){Grohs}, {Howe}, \& {Adams}}]{paperI}
	{Grohs}, E., {Howe}, A.~R., \& {Adams}, F.~C. 2018, \prd, 97, 043003
	
	\bibitem[{{Hall} {et~al.}(2014){Hall}, {Pinner}, \&
		{Ruderman}}]{2014JHEP...12..134H}
	{Hall}, L.~J., {Pinner}, D., \& {Ruderman}, J.~T. 2014, Journal of High Energy
	Physics, 2014, 134
	
	\bibitem[{{Hannestad}(2004)}]{nu_dec_approx_Hannestad}
	{Hannestad}, S. 2004, New Journal of Physics, 6, 108
	
	\bibitem[{{Harnik} {et~al.}(2006){Harnik}, {Kribs}, \& {Perez}}]{harnik}
	{Harnik}, R., {Kribs}, G.~D., \& {Perez}, G. 2006, \prd, 74, 035006
	
	\bibitem[{{Hogan}(2000)}]{hogan}
	{Hogan}, C.~J. 2000, Reviews of Modern Physics, 72, 1149
	
	\bibitem[{{Kane}(2017)}]{kane}
	{Kane}, G. 2017, {Modern Elementary Particle Physics}
	
	\bibitem[{{Kawano}(1992)}]{letsgoeu2}
	{Kawano}, L. 1992, NASA STI/Recon Technical Report, 92, 25163
	
	\bibitem[{{Kolb} \& {Turner}(1990)}]{1990eaun.book.....K}
	{Kolb}, E.~W., \& {Turner}, M.~S. 1990, {The Early Universe.} (Addison-Wesley
	Publishing Co.)
	
	\bibitem[{{Linde}(2017)}]{Linde17}
	{Linde}, A. 2017, Reports on Progress in Physics, 80, 022001
	
	\bibitem[{{Matsuda} {et~al.}(1980){Matsuda}, {Lewis}, \& {Anders}}]{Matsuda80}
	{Matsuda}, J.-I., {Lewis}, R.~S., \& {Anders}, E. 1980, \apjl, 237, L21
	
	\bibitem[{{Mukohyama}(2000)}]{dark_rad:2000}
	{Mukohyama}, S. 2000, Physics Letters B, 473, 241
	
	\bibitem[{{Paxton} {et~al.}(2011){Paxton}, {Bildsten}, {Dotter}, {Herwig},
		{Lesaffre}, \& {Timmes}}]{mesa}
	{Paxton}, B., {Bildsten}, L., {Dotter}, A., {Herwig}, F., {Lesaffre}, P., \&
	{Timmes}, F. 2011, \apjs, 192, 3
	
	\bibitem[{{Planck Collaboration} {et~al.}(2016){Planck Collaboration}, {Ade},
		{Aghanim}, {Arnaud}, {Ashdown}, {Aumont}, {Baccigalupi}, {Banday},
		{Barreiro}, {Bartlett}, \& et~al.}]{PlanckXIII:2015}
	{Planck Collaboration} {et~al.} 2016, \aap, 594, A13
	
	\bibitem[{Rees(1997)}]{rees1997before}
	Rees, M. 1997, Before the Beginning: Our Universe and Others, Helix books
	(Perseus Books)
	
	\bibitem[{{Rees}(2000)}]{reessix}
	{Rees}, M. 2000, {Just six numbers : the deep forces that shape the universe}
	
	\bibitem[{{Sher}(1989)}]{Sher89}
	{Sher}, M. 1989, \physrep, 179, 273
	
	\bibitem[{{Smith} {et~al.}(1993){Smith}, {Kawano}, \& {Malaney}}]{SMK:1993bb}
	{Smith}, M.~S., {Kawano}, L.~H., \& {Malaney}, R.~A. 1993, \apjs, 85, 219
	
	\bibitem[{{Surman} {et~al.}(2013){Surman}, {Mumpower}, {McLaughlin},
		{Sinclair}, {Hix}, \& {Jones}}]{Surman13}
	{Surman}, R.~A., {Mumpower}, M.~R., {McLaughlin}, G.~C., {Sinclair}, R., {Hix},
	W.~R., \& {Jones}, K.~L. 2013, in Capture Gamma-Ray Spectroscopy and Related
	Topics, ed. P.~E. {Garrett} \& {et al.}, 304--313
	
	\bibitem[{{Vassh} {et~al.}(2015){Vassh}, {Grohs}, {Balantekin}, \&
		{Fuller}}]{Vassh_tdecoup}
	{Vassh}, N., {Grohs}, E., {Balantekin}, A.~B., \& {Fuller}, G.~M. 2015, \prd,
	92, 125020
	
	\bibitem[{Wagoner {et~al.}(1967)Wagoner, Fowler, \& Hoyle}]{WFH:1967}
	Wagoner, R.~V., Fowler, W.~A., \& Hoyle, F. 1967, Astrophys.J., 148, 3
	
\end{thebibliography}

% % % % % % % % % % % % % % % % % % % % % % % % % % % % % % % % % % % % % % % % % % % % % % % % % % % % % % % % % % % % % % % % % % %

\end{document}